\documentclass[parskip=true, fontsize = 10pt]{scrartcl} 
\usepackage[margin=1in]{geometry}
\pdfoutput=1 
\usepackage{amsmath}
\usepackage{amssymb}
\usepackage{algorithm,algpseudocode}
\usepackage{algpseudocode}
\usepackage{nicefrac}
\usepackage{pdflscape}
\usepackage{array}
\usepackage{dirtytalk}
\usepackage{placeins}
\usepackage{hyperref}
\DeclareMathOperator*{\argmin}{arg\,min}

\DeclareMathOperator*{\arginf}{arg\,inf}
\usepackage{latexsym}
\usepackage{stmaryrd}
\usepackage{titling}

\usepackage{amsthm}
\usepackage{float}
\usepackage[autostyle=true]{csquotes}
\usepackage{mathtools}
\usepackage{tikz}

\algnewcommand\algorithmicinput{\textbf{Input:}}
\algnewcommand\INPUT{\item[\algorithmicinput]}
\tikzset{ 
treenode/.style = {shape=rectangle, rounded corners, draw, align=center},
  root/.style     = {treenode, font=\large},
  env/.style      = {treenode, font=\normalsize},
  dummy/.style    = {circle,draw}
}
\usetikzlibrary{positioning}
\usetikzlibrary{arrows}
\usetikzlibrary{bending}
\usetikzlibrary{hobby}
\usetikzlibrary{decorations.pathreplacing}
\definecolor{re}{RGB}{165,10,35}
\definecolor{blu}{RGB}{0,40,100}
\definecolor{goal}{RGB}{230,194,0}
\definecolor{ForestGreen}{RGB}{34,139,34}

\newtheorem{prop}{Proposition}

\theoremstyle{definition}
\newtheorem{de}{Definition}

\title{How Morphological Computation shapes Integrated Information in Embodied Agents}
\author{Carlotta Langer$^{1,2 }$ and Nihat Ay$^{1, 2, 3, 4}$
\\
\\
\normalsize{$^{1}$ Hamburg University of Technology, Hamburg, Germany} \\
\normalsize{$^{2}$ Max Planck Institute for Mathematics in the Sciences, Leipzig, Germany} \\
\normalsize{$^{3}$ Leipzig University, Leipzig, Germany} \\
\normalsize{$^{4}$ Santa Fe Institute, Santa Fe, NM, USA}
}
\date{}
\usepackage[backend=biber,bibencoding=utf8,firstinits=true, natbib=true,maxnames=25]{biblatex}
\begin{document}
\vspace{-5cm}
\maketitle
\begin{abstract}
The Integrated Information Theory provides a quantitative approach to consciousness and can be applied to neural networks. An embodied agent controlled by such a network influences and is being influenced by its environment. 
This involves, on the one hand, morphological computation within goal directed action and, on the other hand, integrated information within the controller, the agent’s brain. 
In this article, we combine different methods in order to examine the information flows among and within the body, the brain and the environment of an agent. 
This allows us to relate various information flows to each other. We test this framework in a simple experimental setup. 
There, we calculate the optimal policy for goal-directed behavior based on the “planning as inference” method, in which the information-geometric em-algorithm is used to optimize the likelihood of the goal. 
Morphological computation and integrated information are then calculated with respect to the optimal policies. 
Comparing the dynamics of these measures under changing morphological circumstances highlights the antagonistic relationship between these two concepts. 
The more morphological computation is involved, the less information integration within the brain is required.
In order to determine the influence of the brain on the behavior of the agent it is necessary to additionally measure the information flow to and from the brain. 
\end{abstract}

\section{Introduction}

\subsection{Objective}
An agent that is faced with a task can solve it using solely its brain, its body's interaction with the world, or a combination of those. This article presents a framework to analyze the importance of these different interactions for an embodied agent and therefore aims at advancing the understanding of how embodiment influences the brain and the behavior of an agent. 
To illustrate the idea we discuss the following scenario:

Consider a sailor at sea without any navigational equipment. The sailor has to rely on the information given by the sun or the visible stars in order to determine in which direction to steer. The more complex part of the task is solved by the information processed in the brain of the sailor. On the other hand, a bird equipped with magneto-reception, meaning one that is able to use the magnetic field of the earth to perceive its direction, can rely on this sense and does not need to integrate different sources of information. Here the body of the bird interacts with the environment for the bird to orientate itself. The complexity of the task is met by the morphology of the bird.
Taking this example further we consider a modern boat with a highly developed navigation system. The sailor now only needs to know how to interpret the machines and will therefore have less complex calculations to do. The complexity of the task shifts from the brain and background knowledge of the sailor towards the construction of the navigation system, which receives and integrates different information sources for the sailor to use.

Our objective is to analyze these shifts of complexity. We will do that by quantifying the importance of the information flow in an embodied agent performing a task under different morphological circumstances. 

The importance of the human body for perception of the environment and ourselves is a core idea of the embodied cognition theory, see for example \cite{Wilson} or \cite{GallagherBook}. In \cite{GallagherPaper} Gallagher develops a definition of a human minimal self in the following way:

\begin{quote}
\enquote{Phenomenologically, that is, in terms of how one experiences it, a consciousness of oneself as an
immediate subject of experience, unextended in time. The minimal self almost certainly depends on
brain processes and an ecologically embedded body, but one does not have to know or be aware of
this to have an experience that still counts as a self-experience.}
\end{quote}

Therefore, it is important to understand the influence the
ecologically embedded body has on the brain. Hence, here we
aim at quantifying both, the interaction of the body with the
environment and the information flows inside the body and
the brain, respectively, using the same framework and thereby
relating them to each other. As a first step in that direction we will
analyze simulated artificial agents in a toy example. These agents
have a control architecture, the brain of the agent, consisting of a
neural network. This will provide the basis for future analysis of
more complex agents such as humanoid robots. Ultimately, we
hope to gain insights about human agency, and in particular the
representation of the self.

The setting of our experiment will be presented in Section \ref{sectSetting}. The question we ask is: How is the complexity of solving the task distributed among the different parts of the body, brain and environment? 

The main statements that we will support by our experiments are:
\begin{itemize}
\item[1.] The more the agent can rely on the interaction of its body with the environment to solve a task, the less integrated information in the brain is required. 
\end{itemize}
This antagonistic relationship between integrated information and morphological computation can be observed even in cases in which the controller has no influence on the behavior of the agent. Hence it is necessary to further analyze the information flow in order to fully understand the impact of the controller, the brain of the agent, on the behavior.
\begin{itemize} 
\item[2.] The importance of integrated information in the controller for the behavior of an embodied agent depends additionally on the information flowing to and from the controller. Therefore it is not sufficient to only calculate an integrated information measure.
\end{itemize}
In order to test these statements, we need to develop a theoretical background.

\subsection{Theoretical Background}
We will model the different interactions using the sensori-motor loop, which depicts the connections among the world $W$, the controller $C$, the sensors $S$ and actuators $A$. This will be discussed further in Section \ref{SectWorld}.

Using the sensori-motor loop we are able to define a set of probability distributions reflecting the structure of the information flow of an agent interacting with the world. Now we need to find the probability distributions that describe a behavior that optimizes the likelihood of success. It would be possible to use a learning  or evolutionary algorithm on the agents to find this optimal behavior, but instead we will apply a method called \say{planning as inference}.   

Planning as inference is a technique proposed in \citep{Attias}, in which a goal directed planning task under uncertainty is solved by probabilistic inference tools. This method models the actions an agent can perform as latent variables. These variables are then optimized with respect to a goal variable using the em-algorithm, an information geometric algorithm that is guaranteed to converge, as proven in \citep{EMvsem}. This algorithm might result in local minima depending on the input distribution, which allows us to analyze different kinds of agents and strategies that lead to a similar probability of success.
This course of action has the advantage that we can directly calculate the optimal policies without having to train the agents. We will describe this method in the context of our experimental setup in further detail in Section \ref{planningasinference}.

Having calculated the distributions that describe the optimal behavior, we apply various information theoretic measures to quantify the strength of the different connections. The measures we are going to discuss are defined by minimizing the KL-divergence between the original distribution and the set of split distributions. The split distributions lack the information flow that we want to measure.
Following this concept we are able to quantify the strength of the different information flows, which  leads to measures that can be seen in the context of integrated information and morphological computation. We will further define four additional measures that together quantify all the connections among the controller, sensors and actuators. These are defined in Section \ref{measures}.

Using information theoretic measures to quantify the information flow in an embodied agent is a natural approach, since we could perceive the different parts of the system as communicating with each other. Surely the world does not actively send information to the controller, but the controller still receives information about the world through the sensors. Therefore there have been different studies analyzing acting agents by using information theoretic measures.  In \citep{MaximizeInformaionFlow} maximizing the information flow through the whole system is used as a learning objective.  Furthermore, in \citep{InfTheoret} the authors use the concepts of information and entropy to define conditions under which a system is perfectly controllable or observable. 
Emphasizing the importance of the sensory input, entropy and mutual information are utilized in \citep{Sporns2004}  to analyze how an agent actively structures its sensory input. Moreover, the authors of  \citep{QuantInformStructure} also include the structure of the motor data in their analysis. The last two cited articles additionally discuss two measures regarding the amount of information and the complexity of its integration. These concepts are also important in the context of Integrated Information Theory.

Integrated Information Theory (IIT) proposed by Tononi aims at measuring the amount and quality of consciousness. This theory went through multiple phases of development starting as a measure for brain complexity \citep{braincomplexity} and then evolved through different iterations \citep{3}, \citep{6}, towards a broad theory of consciousness \citep{12}.
The two key concepts that are present in all versions of IIT are \say{Information} and \say{Integration}. Information refers to the number of states a system can be in and Integration describes the amount to which the information is integrated among the different parts of it. Measures for integrated information differ depending on the version of the theory they are referring to and on the framework they are defined in.
We discussed a branch of these measures building on information geometry in \citep{CII}. In this article we will use the measure that we propose in \cite{CII} in the case of a known environment, as defined in Section 2.3.1. As
advocated by the authors of \cite{mediano2021integrated} we will treat the integrated information measure as a complexity measure and therefore as a way to quantify the relevant information flow in the controller.

Another general feature of all IIT measures so far is that they focus solely on the brain, meaning on the controller in the case of an artificial agent. 
Therefore we want to embed these measures into the sensori-motor loop and analyze their behavior in relation to the dynamics of the body and environment.
Although the measures are only focusing on the controller, there have been simulated experiments with evolving embodied agents, interacting with their environment, in the context of IIT. In \citep{9a} the authors measure the values for integrated information for simulated evolving artificial agents in a maze and conclude that integrated information grows with the fitness of the agents.  
Increasing the complexity of the environment leads in \citep{AnimatsIncreasing} to the conclusion that integrated information needs to increase in order to capture a more complex environment. In \citep{13} the authors go one step further and conclude from experiments with elementary cellular automata and adaptive logic-gate networks that a high integrated information value increases the likelihood of a rich dynamical behavior. All of these examples focus on the measures in the controller in order to analyze what kind of cause-effect structure makes a difference intrinsically. Since we are interested in an embodied agent solving a task, we want to emphasize the importance of the interaction of the agent's body with the world and additionally measure this interaction explicitly. This leads us to the concept of morphological computation.

Morphological computation is the reduction of computational complexity for the controller resulting from the interaction between the body and the world, as described in \citep{MorphologicalIntelligence}. There are different ways in which the body can lift the burden of the brain, as discussed in \citep{MullerMorph}. An example for morphological computation is the bird using its magneto-reception mentioned earlier in the introduction. Another case of morphological computation would be a human grabbing a fragile object compared to a robotic metal hand. The soft tissue of the human hands allows us to be less precise in the calculation of the pressure that we apply. The robot needs to perform more difficult computations and will therefore most likely have a higher integrated information. Does this mean that our experience of this task is less conscious than the experience of the robot? Here we want to take a step back from the abstract concept of consciousness and instead examine the complexity of the tasks. Even though the interactions are not fully controlled by the human brain, the soft skin of the human hand interacts with the object in a more complicated manner than the robot's hand.
In this article we want to analyze how the complexity of solving a task is met by the different information flows among the brain, body and environment. In \citep{MappSens}, the authors find that the information flow in the agent can be affected by changes in the body's morphology. Examining this phenomenon further we will observe shifts in the importance of the information flows depending on the morphology of the body, which directly changes the complexity of the environment for the agent.  

Furthermore, we will define two additional groups of agents. For the agents of the first group all the information has to go through the controller, while the controller has no impact on the action for the agents in the second group. These cases demonstrate once more that the antagonistic behavior of morphological computation and integrated information exists regardless of the behavior of the agents.  
The results of our experiments are presented in Section \ref{Results}. 

\section{Materials and Methods}

\subsection{Setting} \label{sectSetting}
\begin{figure}
\begin{center}
\includegraphics[width = 0.9\textwidth]{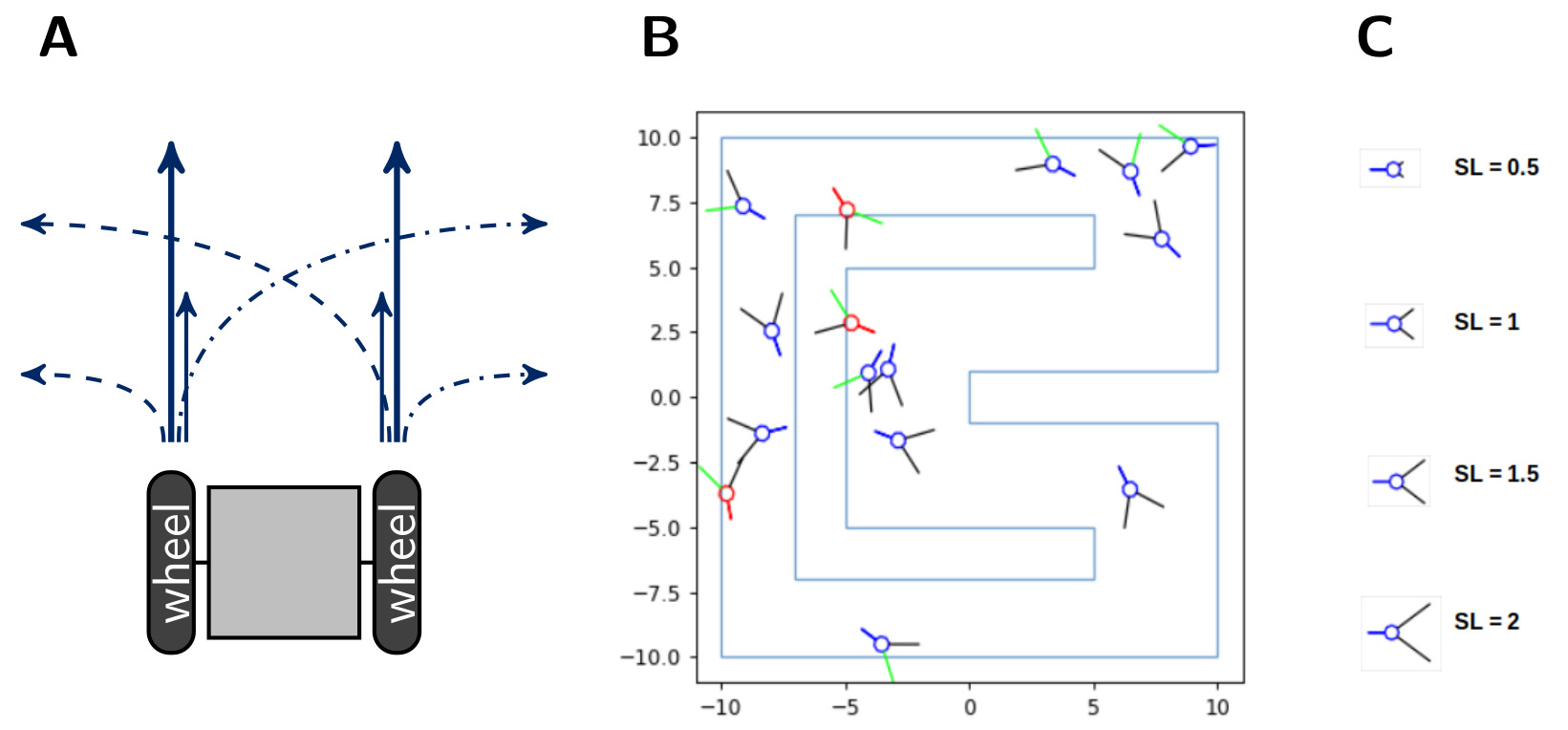}
\end{center}
\caption{(A) A sketch of a two-wheeled robot and its four different types of movement. (B) The racetrack the agents have to survive in and  (C) the different sensor lengths, named SL, on the right.}  \label{Fig1}
\end{figure}

In order to analyze the information flow of an acting agent, we examine the following simple setting.
The agents are idealized models of a two-wheeled robot depicted in Figure \ref{Fig1} (A). Each wheel can spin either fast or slow, hence the agents have four different movements and are unable to stop. If both wheels spin fast, then the agent moves 0.6 units of length and if they both spin slow, then the agent moves 0.2. In case
of one fast and one slow wheel the agent makes a turn of approx.~10 degrees with a speed of 0.4. The code
of the movement of the agents and a video of 5 agents performing random movements can be found in
\cite{GitHub}.
The agent's body consists of a blue circle and a blue line marking the back of the agent, depicted in Figure \ref{Fig1} (B). The two black lines are  binary sensors that only detect whether they touch an obstacle or not, without reporting the exact distance to it. If a sensor touches a wall it turns green and if the body of the agent touches a wall it turns red.
 
Consider a racetrack as shown in Figure \ref{Fig1} (B). The agents die as soon as their bodies touch a wall. Hence the goal for the agents is to stay alive. The design and implementation of the agents and the  racetrack is due to Virgo, \citep{Nathaniel}.  Although we depicted more than one agent in the environment, these agents do not influence each other.

Additionally, we want to manipulate the amount of potential morphological computation for the agents.
There exist different concepts referred to as morphological computation, as thoroughly examined in \citep{MullerMorph}, where the authors distinguish between three different categories. These are (1) Morphology facilitating control,
(2) Morphology facilitating perception and 
(3) proper Morphological computation.
The notion we will use belongs to the second category and is called \say{pre-processing} in  \citep{MorphologicalIntelligence}. How well agents perceive their environment can heavily influence the complexity of the task they are facing. One example is the design of the compound eyes of flies, which has been analyzed and used for building an obstacle avoiding robot in \cite{FlyEye}. Therefore manipulating the
qualities of the sensors directly influences the agent’s perception and consequently the amount of necessary
computation in the controller.  Hence changing the length of the sensors influences the agent's ability for morphological computation.
We will therefore vary the length of the sensors from 0.5 to 2.75. Four different sensor lengths are
depicted in Figure \ref{Fig1} (C).

The strategies the agents should use will be calculated by applying the concept of planning as inference as discussed in Section \ref{planningasinference}. Utilizing this method we are able to directly determine the optimal behaviors without having to train any agents. 

Before we discuss this further, we will first present the control architecture of the agents in the next section.

\subsubsection{The Agents} \label{Sect:Agents}

\begin{figure}
\centering
\includegraphics[width = 0.75\textwidth]{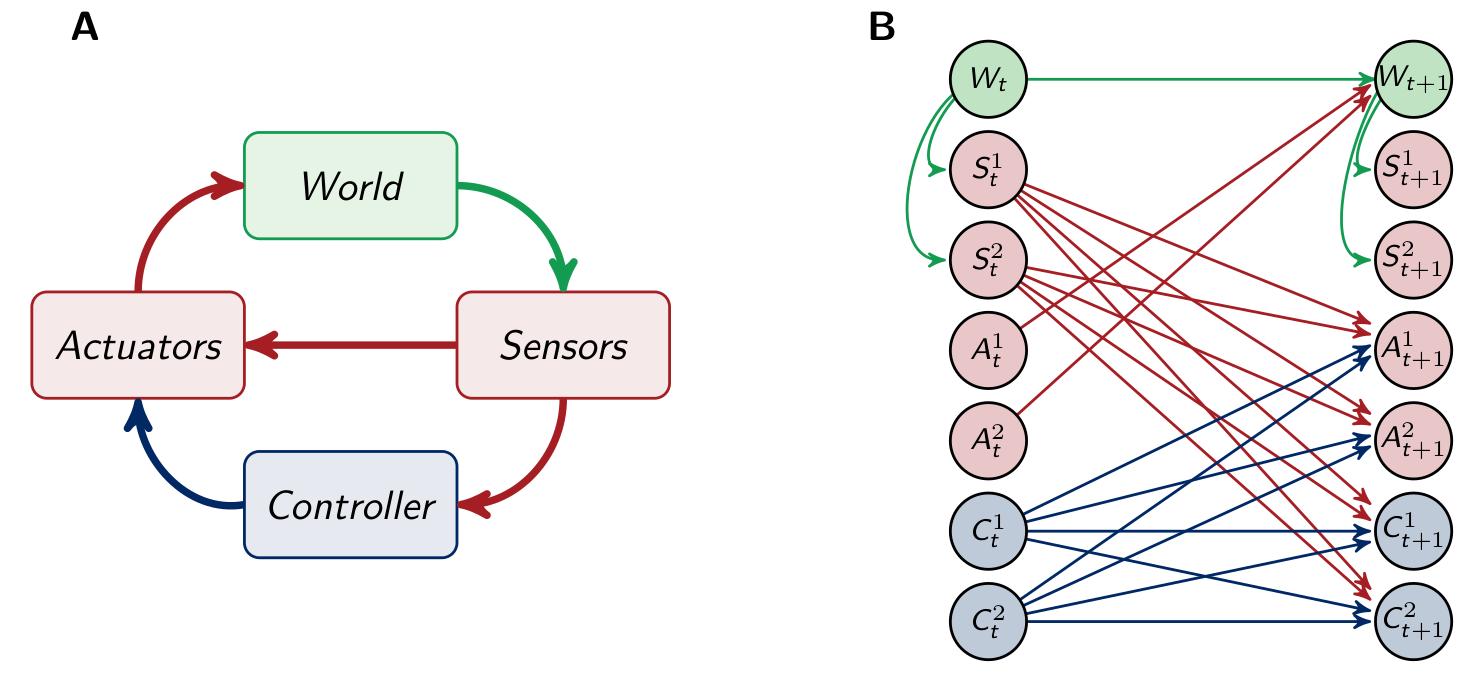}
\caption{(A) The sensori-motor loop and (B) the architecture of the agents.} \label{Fig2} 
\end{figure}

We model the whole system by using the sensori-motor loop as depicted in Figure \ref{Fig2} (A). There the information about the world is received by be the sensors, which send their information to the controller and directly to the actuators. This direct connection between the sensors and the actuators enables the agent to have a response to certain stimuli, without the need for integrating the information in the controller. The
controller processes the information from the sensors and also influences the actuators, which in turn have an effect on the world. The sensori-motor loop, also called action-perception circle, has been analyzed and discussed in, for example, \cite{Klyubin}, \cite{Ay2014} and \cite{Ay2015b}.

Unfolding the connections among the different parts of the agent and its environment for one timestep
leads to the depiction in Figure \ref{Fig2} (B).
The architecture of the agents is the following one. 
The agents have two sensor $S_{t}^{1}, S_{t}^{2}$, two controller $C_{t}^{1}, C_{t}^{2}$ and two actuator nodes $A_{t}^{1}, A^{2}_{2}$. The sensors and controllers send their information to the actuators and controllers in the next point in time. The sensors are only influenced by the world $W$ and the world is only affected by the actuators and the last world state as depicted in Figure \ref{Fig2}. 

To simplify we only draw one node for each $S,A$ and $C$ in the following graphs. 

The behavior of the agents is governed by a probabilistic law, which can be modeled as the following discrete multivariate time-homogeneous Markov process
\begin{equation*}
(X_{t})_{t \in \mathbb{N}} = (W_{t},S_{t},A_{t},C_{t})_{t \in \mathbb{N}}
\end{equation*}
with the state space $\mathcal{X} = \mathcal{W} \times \mathcal{S} \times \mathcal{A} \times \mathcal{C}$ and the distribution 
\begin{equation*}
\begin{aligned}
P(x_{0}, \dots, x_{t+1}) &= P(x_{0}) \prod\limits_{i=1}^{t+1} P(x_{t} \vert x_{t-1}) \\
P(x_{t+1} \vert x_{t}) &=  P(w_{t+1} \vert w_{t}, a_{t}) \prod\limits_{k} P(s^{k}_{t+1} \vert w_{t+1}) \prod\limits_{i} P(a^{i}_{t+1} \vert s_{t},c_{t}) \prod\limits_{j} P(c^{j}_{t+1} \vert s_{t},c_{t}).
\end{aligned}
\end{equation*}

The corresponding directed acyclic graph is depicted in Figure \ref{Fig3} (A). See \citep{graphModels} for more information on the relationship between graphs and graphical models. Throughout this article we will assume that the distributions on $\mathcal{X}$ are strictly positive.

In the next section we will take a closer look at the role of the environment.

\subsubsection{The Environment} \label{SectWorld}

\begin{figure}[ht]
\centering
\includegraphics[width = 0.55\textwidth]{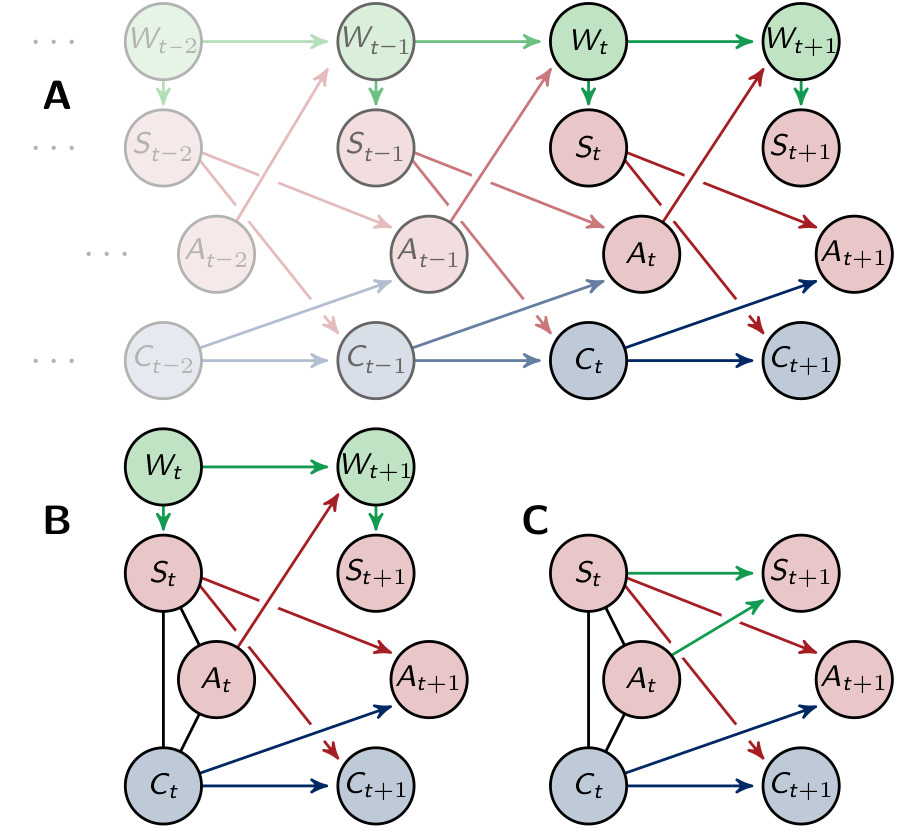}
\caption{(A) Graphical representation of the Markov process $(X_{t})_{t \in \mathcal{N}}$. (B) Graphical representation of one timestep and (C) the marginalized graph.} \label{Fig3} 
\end{figure}

The Markov process defined above describes the interactions between the agent and its environment in terms of a joint distribution. Note that the distributions discussed in this section determine the information flow in the system. The optimization of this flow will require a planning process which we are going to address in the next section.
Since the agent has only access to the world through the sensors, we replace
\begin{equation*} P(w_{t+1} \vert w_{t}, a_{t}) \prod\limits_{k} P(s^{k}_{t+1} \vert w_{t+1})
\end{equation*} 
using only information intrinsically known to the agent. In order to do that, we will look closer at one step in time $P(x_{t}, x_{t+1}) = P(x_{t}) P(x_{t+1} \vert x_{t})$. Reducing the focus to one step in time means that we need to define
an initial distribution that takes into account the past of the agent. In Figure \ref{Fig3} (A) we see that the sensors
$S_{t}$, actuators $A_{t}$ and controller nodes $C_{t}$ are conditionally independent given the past, but marginalizing to
the point in time $t$ leads to additional connections. More precisely marginalizing to one timestep results in undirected edges between $S_{t}$ , $A_{t}$ and $C_{t}$ . Here we will assume that the environment only influences the sensors, even in the  graph marginalized to one timestep as depicted in Figure  \ref{Fig3} (B). We will then sum over $w_{t},w_{t+1} \in \mathcal{W}$ in order to get a Markov process that only depends on the variables known to the agent.

\begin{prop} \label{SamplingEnv}
Marginalizing the distribution, that corresponds to the graph (B) in Figure \ref{Fig3}, that is
\begin{equation*}
P(x_{t},x_{t+1}) = P(w_{t}) \cdot P(s_{t}, a_{t}, c_{t} \vert w_{t}) \cdot P(w_{t+1} \vert w_{t}, a_{t}) \prod\limits_{k} P(s^{k}_{t+1} \vert w_{t+1}) \prod\limits_{i} P(a_{t+1}^{i} \vert s_{t},c_{t}) \prod\limits_{j} P(c^{j}_{t+1} \vert s_{t},c_{t})
\end{equation*} 
over $(w_{t}, w_{t+1}) \in \mathcal{W} \times
\mathcal{W}$  leads to the following Markov process
\begin{equation*}
P(s_{t}, a_{t}, c_{t}, s_{t+1}, a_{t+1}, c_{t+1}) = P(s_{t}, a_{t}, c_{t}) \cdot \prod\limits_{i} P(a^{i}_{t+1} \vert s_{t},c_{t}) \prod\limits_{j} P(c^{j}_{t+1} \vert s_{t},c_{t}) \cdot  P(s_{t+1} \vert s_{t}, a_{t}).
\end{equation*} 
\end{prop}
The proof can be found in the Appendix.

The new process describes the behavior of the environment with information known by the agent and is shown in Figure \ref{Fig3} (C).
A similar distribution is also used in \citep{QuantifyingMorphComp} in Section 3.3.1.~and in \citep{MorphologicalIntelligence}. There it is derived by taking $P(S_{t+1} \vert S_{t})$ as the intrinsically available information of $P(W_{t+1} \vert W_{t})$.

We sample this distribution $\tilde{P}(S_{t+1},S_{t},A_{t})$ for every sensor length, by storing 20.000.000 sensor and motor values for agents starting in a random place in the arena, performing arbitrary movements. We denote all the sampled and therefore fixed distributions by $\tilde{P}$.

Since we are now able to define a set of distributions that describe the interaction between the agent and the world according to the sensori-motor loop, we will present the method to find the optimal behavior in the next section.

\subsection{Optimizing the Behavior} \label{planningasinference}

In order to calculate the optimal behavior of the agents, we will use the concept of planning as inference. This was originally proposed by Attias in \citep{Attias} and further developed by Toussaint and collegues in \citep{Toussaint06}, \citep{Toussaint09} and  \citep{Toussaint12} as a theory of planning under uncertainty. There the conditional distribution describing the action of the agent is considered to be a hidden variable that has to be optimized.
This is done by using the EM-algorithm, which is equivalent to the information theoretic em-algorithm in this case. We describe the optimization using the em-algorithm, because of its intuitive geometric nature. More details can be found in the Appendix. The em-algorithm is well known and was proposed in 1984 in \citep{Csiszar}, further discussed in \citep{EMvsem} and \citep{EM}. The resulting distribution maximizes the likelihood of achieving the predefined goal, but might be a local optimum depending on the initial distribution. Normally this is a disadvantage, but in our setting it allows us to analyze various strategies by using different initial distributions. 

\begin{figure}[ht]
\centering
\includegraphics[width = 0.65\textwidth]{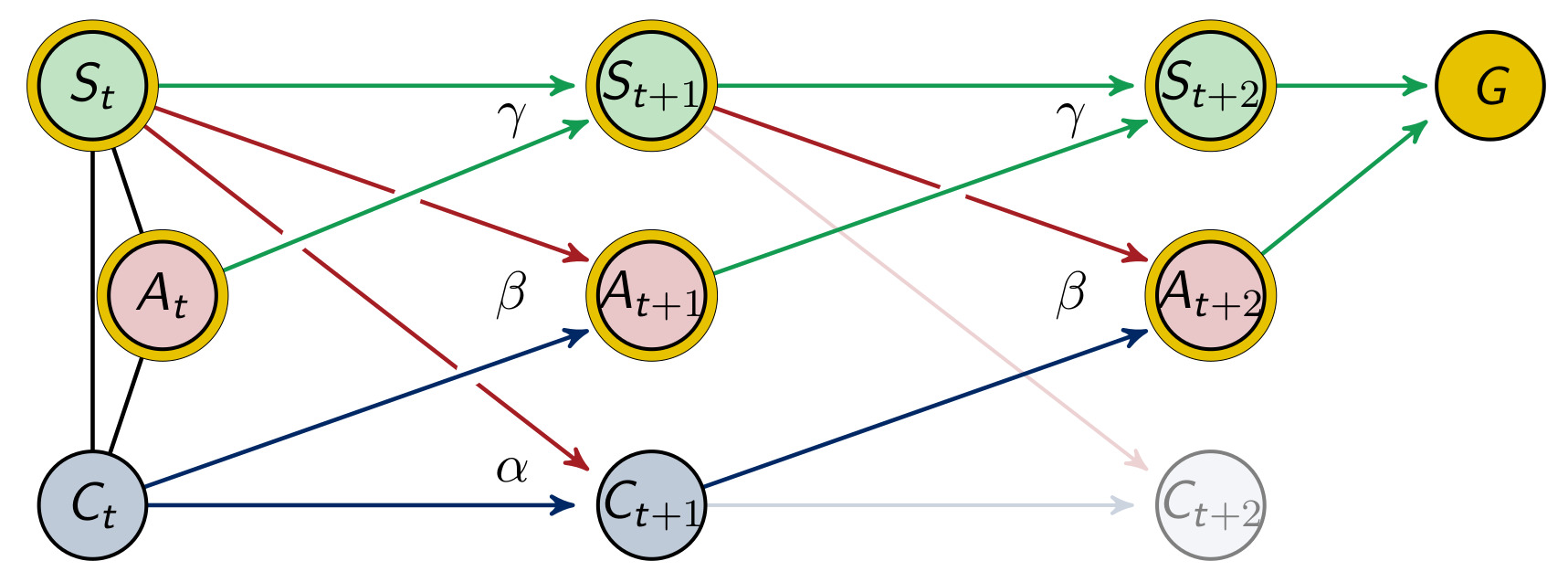}
\caption{Graphical representation of two timesteps. } \label{Fig4}
\end{figure}

The goal of the agents in our example is to maximize the probability of being alive after the next two movements. To make at least two steps is necessary since we want the connection between $C_{t}$ and $C_{t+1}$ to have an impact on the outcome. This can be seen in Figure \ref{Fig4}.

We will denote the goal variable by $G$ with the state space $\mathcal{G} = \{0,1\}$,  where $P(g_{1}):=P(g=1)$ refers to the probability of the agent to be alive. Since the agent moves twice, this distribution depends on the states of the last three sensor and motor states
\begin{equation*}
\tilde{P}(G \vert S_{t+2}, S_{t+1}, S_{t}, A_{t+2}, A_{t+1}, A_{t}).
\end{equation*}
The node $G$ depends on the ones that are marked with a golden circle in Figure \ref{Fig4}. We sampled this distribution for every sensor length, as described in the previous section in the context of $\tilde{P}(S_{t+1}, S_{t}, A_{t})$. 

The architecture of the agents considered in this article was discussed in the last sections. There we outlined how we sample the distribution $\gamma = \tilde{P}(S_{t+1} \vert S_{t}, A_{t})$ that describes the influence the agent has on itself through the world. The distributions influencing the behavior of the agents are 
\begin{equation*}
\beta =  P(A_{t+1} \vert S_{t},C_{t}) \quad \quad \text{   and   } \quad \quad \alpha =  P(C_{t+1} \vert S_{t}, C_{t}).
\end{equation*} 
Hence we will treat $(A_{t+1}, C_{t+1})$ as hidden variables and optimize their distributions with respect to the goal. We denote these distributions by $\alpha, \beta$ and $\gamma$ in order to emphasize that the process is time-homogeneous, meaning that $P(A_{t+1} \vert S_{t}, C_{t}) = P(A_{t+2} \vert S_{t+1}, C_{t+1})$, $ P(S_{t+1} \vert S_{t}, A_{t}) =  P(S_{t+2} \vert S_{t+1}, A_{t+1})$ and  $ P(C_{t+1} \vert S_{t}, C_{t}) =  P(C_{t+2} \vert S_{t+1}, C_{t+1})$ as indicated in Figure \ref{Fig4}. Note that the above mentioned homogeneity does not imply stationarity.

It remains to define the initial distribution  $P(S_{t},C_{t},A_{t})$.In the original planning as inference framework an action sequence is selected conditioned on the final goal state and an initial observation, as described in
\cite{Attias}. Here, we do not want to restrict the agents to an initial observation $S_{t}$. Instead we first write the initial distribution in the following form
\begin{equation*}
P(s_{t},c_{t},a_{t}) = P(c_{t} \vert a_{t}, s_{t}) P(s_{t} \vert a_{t}) P(a_{t}).
\end{equation*}
Using the sampled distribution $\tilde{P}(S_{t+1}, S_{t}, A_{t})$, we are able to calculate $ \tilde{P}(S_{t} \vert A_{t})$ and set $P(s_{t} \vert a_{t}) = \tilde{P}(s_{t} \vert a_{t})$. The remaining distributions  $P(c_{t} \vert a_{t}, s_{t})$ and $P(a_{t})$ are also treated as variables and optimized using the em-algorithm. This approach leads to the optimal starting conditions for the agents. 
The details of the optimization are described in the Appendix.

\subsection{Measures of the Information Flow} \label{measures}

In this section we will define the different measures. These are information theoretic measures that use the KL-divergence to calculate the difference between the original distribution and a split distribution. This split distribution is the one that is closest to the original distribution without having the connection that we want to measure. 

\begin{de}[Measure $\Psi$]
Let $M \subset P^{\circ}(\mathcal{Z})$ be a set of probability distributions corresponding to a split system.
Then we define the measure $\Psi$, by minimizing the KL-divergence between $M$ and the full distribution $P$ to quantify the strength of the connections missing in the split system
\begin{equation*}
\Psi = \inf\limits_{Q \in M} D(P \parallel Q) = \sum\limits_{z} P(z) \, log \, \dfrac{P(z)}{Q(z)}.
\end{equation*} 
Note that this measure depends on $M$, the set of split distributions.
\end{de}

Every discussed measure has a closed form solution and can be written in the form of sums of conditional mutual information terms. 

\begin{de}[Conditional Mutual Information] \label{MI}
Let $(Z_{1},Z_{2}, Z_{3})$ be a random vector on $\mathcal{Z} = \mathcal{Z}_{1} \times \mathcal{Z}_{2} \times \mathcal{Z}_{3}$ with the distribution $P$.
The conditional mutual information of the random variables $Z_{1}$ and $ Z_{2}$ given $Z_{3}$  is defined as
\begin{equation*}
\begin{aligned}
I(Z_{1} ; Z_{2} \vert Z_{3}) &= \sum\limits_{z_{1} \in \mathcal{Z}_{1}}  \sum\limits_{z_{2} \in \mathcal{Z}_{2}}  \sum\limits_{z_{3} \in \mathcal{Z}_{3}} P(z_{1}, z_{2},z_{3}) \, log \, \left( \dfrac{P(z_{1}, z_{2} \vert z_{3})}{P(z_{1} \vert z_{3}) P(z_{2} \vert z_{3})} \right) \\
&= \sum\limits_{z_{1} \in \mathcal{Z}_{1}}  \sum\limits_{z_{2} \in \mathcal{Z}_{2}}  \sum\limits_{z_{3} \in \mathcal{Z}_{3}} P(z_{1}, z_{2},z_{3}) \, log \, \left( \dfrac{P(z_{1} \vert z_{2}, z_{3})}{P(z_{1} \vert z_{3})} \right) .
\end{aligned}
\end{equation*}

\end{de}

If $I(Z_{1} ; Z_{2} \vert Z_{3}) = 0$, then $Z_{1}$ is independent of $Z_{2}$ given $Z_{3}$. Therefore this quantifies the connection between $Z_{1}$ and $Z_{2}$, while $Z_{3}$ is fixed. 
Additionally, we emphasize this by marking the respective connection quantified by the measure in a graph as a dashed connection. To simplify the figures we only depict one timestep, but the connections between $(Y_{t+1}, Y_{t+2})$ are the same as the connections between $(Y_{t}, Y_{t+1})$. 

The base of the logarithms in the definitions above is 2, hence the unit of all the measures defined below is bits.

Although these measures were originally defined for only one timestep, we will introduce them directly tailored to our setting with two timesteps. 
  
\subsubsection{Integrated Information and Morphological Computation}

The two measures discussed in this section each quantify the information flow among the same type of node in different points in time. Integrated information only considers the nodes inside the controller and therefore measures the information flow inside the agent, while morphological computation is concerned with the exterior perspective and measures the information flow between the sensors.

\begin{figure}[ht]
\centering
\includegraphics[width=0.6 \textwidth]{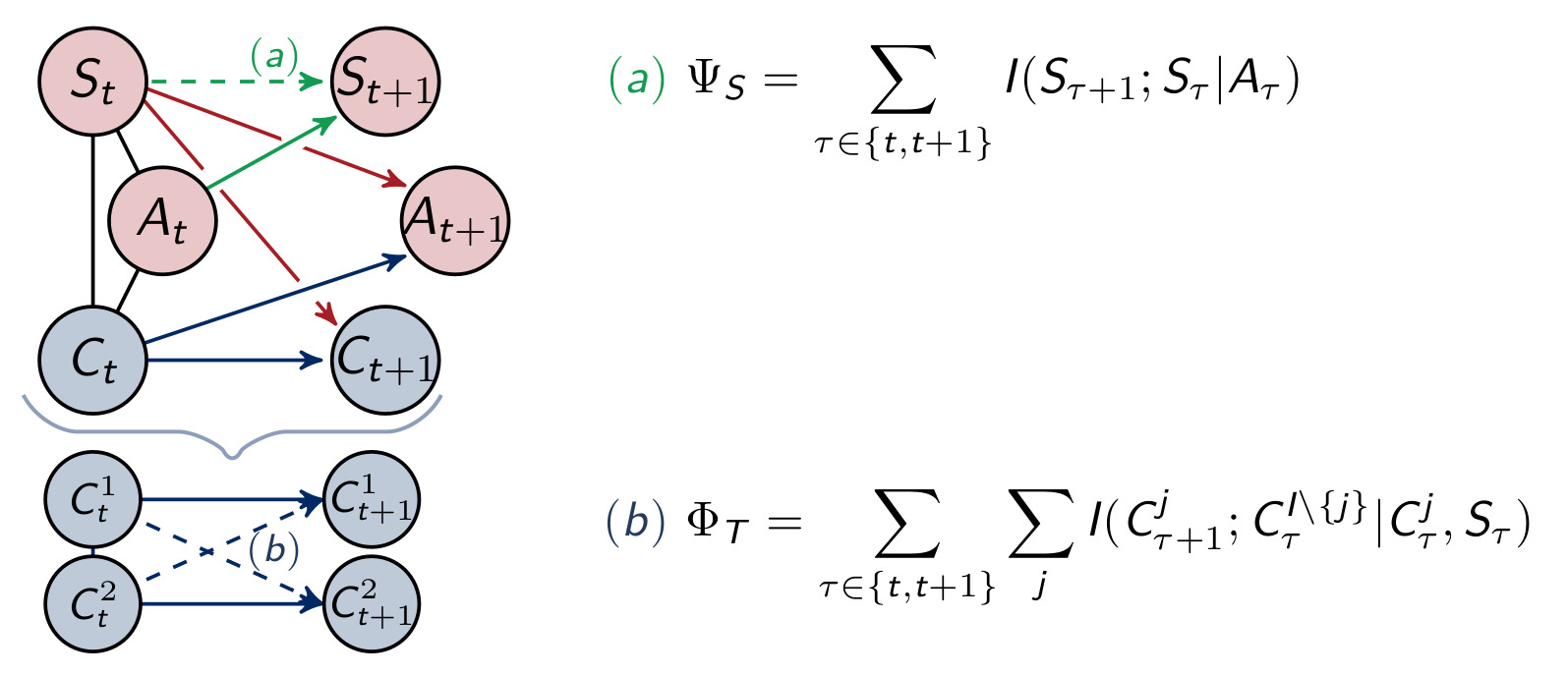}
\caption{Calculation of the measures for morphological computation (a) and  integrated information (b).} \label{Fig5}
\end{figure} 

\paragraph{Integrated Information}
 
The measure $\Phi_{T}$ restricts itself to the controller nodes and can be seen in the context of the Integrated Information Theory of consciousness \citep{Tononi}. 
This theory was discussed in the introduction. 
It aims at measuring the strength of the connections among different nodes across different points in time, in other words, the connections that integrate the information. 
Since every influence on $C_{t+1}$ is known in our setting, we are able to use the measure $\Phi_{T}$ proposed in \citep{CII}. 
This measure is defined in the following way
\begin{equation*}
\Phi_{T} = \sum\limits_{\tau \in \{t, t+1\}} \sum\limits_{j} I(C^{j}_{\tau +1}; C^{I \setminus \{ j\} }_{ \tau} \vert C_{\tau }^{j}, S_{\tau})  
\end{equation*}
and depicted as (b) in Figure \ref{Fig5}.  In the definition above, $I(C^{j}_{t+1}; C^{I \setminus \{j\}}_{t} \vert C_{t}^{j}, S_{t})$ denotes the conditional mutual information, described in Definition \ref{MI}, and $I \setminus \{j\}$ is the set of indices of controller nodes without $j$. For two controller nodes and $j=2$ this would be $\{1,2\} \setminus \{ 2\} = \{1\}$. Hence $\Phi_{T}$ measures the connections between $C^{i}_{t}$ and $C^{j}_{j+1}$ with $i,j \in \{1,2\}$ and $i \neq j$.

A proof of the closed form solution can be found in \citep{CII}. All the following measures can be proven in a similar way.

\paragraph{Morphological Computation}

In \citep{MorphologicalIntelligence} morphological computation was referred to as morphological intelligence and characterized in Definition 1.1.~as follows
\begin{quote}
\say{Morphological Intelligence is the
reduction of computational cost for the brain (or controller) resulting from the
exploitation of the morphology and its interaction with the environment.}
\end{quote}
There exists a variety of measures for morphological computation.
The distribution $\tilde{P}(S_{t+1} \vert S_{t}, A_{t})$ describes the influence the agent has on itself through the environment. Hence this distribution is dependent on the environment and the morphology of the agent. The interplay between environment and body is influenced by the length of the sensors. 

In \citep{QuantifyingMorphComp} the authors define the following measure for morphological computation, which depends on $\tilde{P}(S_{t+1} \vert S_{t}, A_{t})$. It quantifies the strength of the influence of the past sensory input on the next sensory input given the last action as
\begin{equation*}
\Psi_{S} =  \sum\limits_{\tau \in \{t, t+1\}}  I(S_{\tau+1} ; S_{\tau} \vert A_{\tau}) 
\end{equation*}
which corresponds to $ASOC_{W}$ defined in \citep{MorphologicalIntelligence} in Definition 3.1.3. There the author compares the different measures numerically and concludes in the chapter 4.9 that the measure following the approach of $\Psi_{S}$, but defined directly on the world states, has advantages over other formulations and is therefore the recommended one. We will follow this reasoning and consider $\Psi_{S}$ to be the measure of morphological computation.

\subsubsection{Measures for Information Flows between different Types of Nodes }

\begin{figure}[ht]
\centering
\includegraphics[width = 0.9 \textwidth]{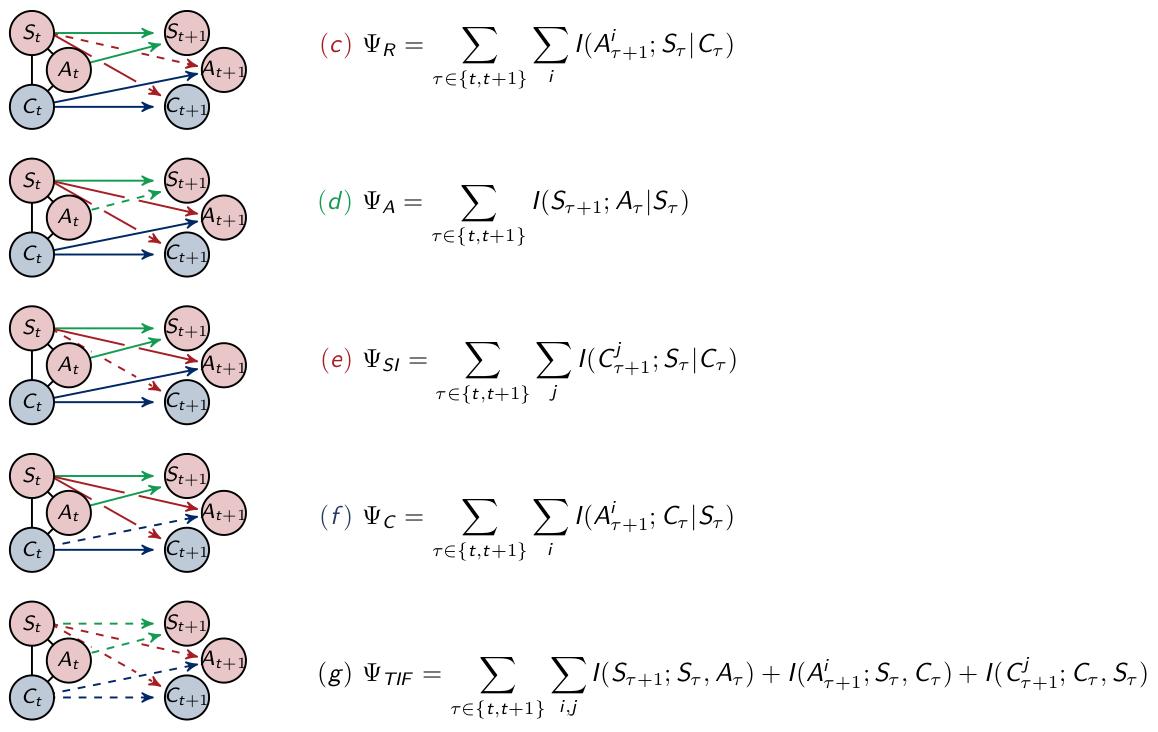}
\caption{Calculation of the measures for (c) reactive control, (d) action effect, (e) sensory information,
(f) control and (g) total information flow.} \label{Fig6}
\end{figure}

We will observe that the measures for integrated information and morphological computation behave antagonistically. This, however, does not lead to a definitive conclusion about how much of the behavior of the agent is determined by the controller. Intuitively, it might be the case, that the agent acts regardless of all the information integrated in the controller. In order to understand the influences leading to the actions of the agents, we will present four additional measures for the four remaining connections in the graph. These are depicted in Figure \ref{Fig6}. 

\paragraph{Reactive Control}

Reactive control describes a direct stimuli response, meaning that the sensors send their unprocessed information directly to the actuators. We are measuring this by the value $\Psi_{R}$. The corresponding split distribution results from removing the connection between $S_{t}$ and $A_{t+1}$
\begin{equation*}
\Psi_{R}  = \sum\limits_{\tau \in \{t, t+1\}}  \sum\limits_{i} I(A^{i}_{\tau +1} ; S_{\tau} \vert C_{\tau}).
\end{equation*}

\paragraph{Action Effect}
We are able to quantify the effect of the action on the next sensory state by calculating
\begin{equation*}
\Psi_{A} = \sum\limits_{\tau \in \{t, t+1\}}  I(S_{\tau+1} ; A_{\tau} \vert S_{\tau}).
\end{equation*}
This measures the amount of control an agent has. Hence in \citep{QuantifyingMorphComp} this measure was normalized and inverted in order to quantify morphological computation. The differences between this approach and $\Psi_{S}$ are further discussed in Section 4.9 in \citep{MorphologicalIntelligence}.

\paragraph{Sensory Information}

The commands the controller sends to the actuators should be based on the information received from the sensors. 
Therefore we will additionally calculate the strength of the information flow from the sensor to the controller nodes. 
The smaller this value is, the more likely it is that the controller converged to a general strategy and performs this blindly without including the information from the sensors. 
We will call this \say{sensory information}, $\Psi_{SI}$. 
\begin{equation*}
\Psi_{SI} =  \sum\limits_{\tau \in \{t, t+1\}}   \sum\limits_{j} I(C_{\tau +1}^{j}; S_{\tau} \vert C_{\tau}) 
\end{equation*}

\paragraph{Control}

Since we are looking at an embodied agent, we additionally want to measure how much of the information processed in the controller has an actual impact on the behavior of the agent. 
We will term the measure quantifying the strength of the impact of the controller on the actuators \say{control}, $\Psi_{C}$. 
\begin{equation*}
\Psi_{C} =  \sum\limits_{\tau \in \{t, t+1\}}  \sum\limits_{i} I(A_{\tau +1}^{i}; C_{\tau} \vert S_{\tau}) 
\end{equation*} 

\paragraph{Total Information Flow}

The last measure quantifies the total information flow, $\Psi_{TIF}$. In this case two points in time are independent of each other in the split system, as depicted in Figure \ref{Fig6},  

\begin{equation*}
\Psi_{TIF} =  \sum\limits_{\tau \in \{t, t+1\}}  \sum\limits_{i,j} I(S_{\tau+1}; S_{\tau}, A_{\tau}) +I(A_{\tau+1}^{i}; S_{\tau}, C_{\tau}) + I(C_{\tau+1}^{j}; C_{\tau}, S_{\tau}).
\end{equation*} 

The total information flow is an upper bound for all the other measures defined in the previous sections.

\section{Results} \label{Results}
In this section we will present the results of our experiments.
The length of the sensors are varied from 0.5 to 2.75 in steps of 0.25. We took 100 random input distributions $\bar{P}$. Each time the algorithm takes at least 1000 iteration steps and stops when the difference between the likelihood of the goal is smaller than $1*10^{-5}$. 

\subsection{Fully coupled Agents}

\begin{figure}[ht]
\centering
\includegraphics[width = 1\textwidth]{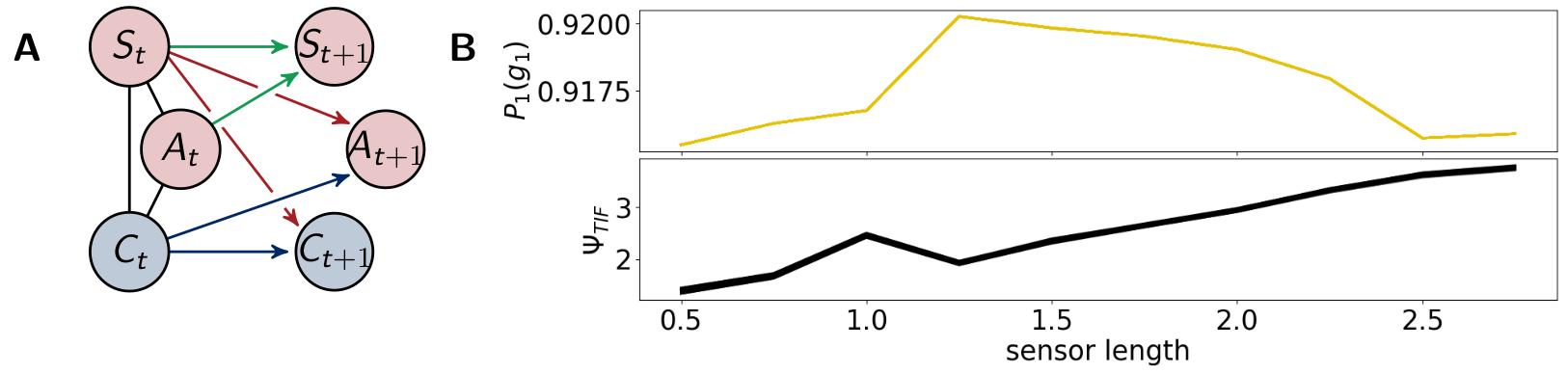}
\caption{(A) The architecture of the fully coupled agents and (B) the probability of survival (top) and the
total information flow $\Psi_{TIF}$ (bottom).} \label{Fig7}
\end{figure}

The architecture of the fully coupled agents are the ones described in Section \ref{Sect:Agents} as shown in Figure \ref{Fig7} on the left. We will refer to the optimized distribution of a natural agent by $P_{1}$, hence $P_1(g_1)$ is the probability with which the agents survive. This value is depicted in Figure \ref{Fig7} on the right.
The agents perform best between a sensor length of 1.25 and 2.25. If the sensors are too long or too short their information is not useful to assure the survival of the agents.

The total information flow $\Psi_{TIF}$ in Figure \ref{Fig7} on the bottom right exhibits an almost monotonic increase, except for a local maximum at a sensor length of 1. We will discuss this sensor length below in the context of Figure \ref{Fig9}.

\begin{figure}[h]
\centering
\includegraphics[width=0.8\textwidth]{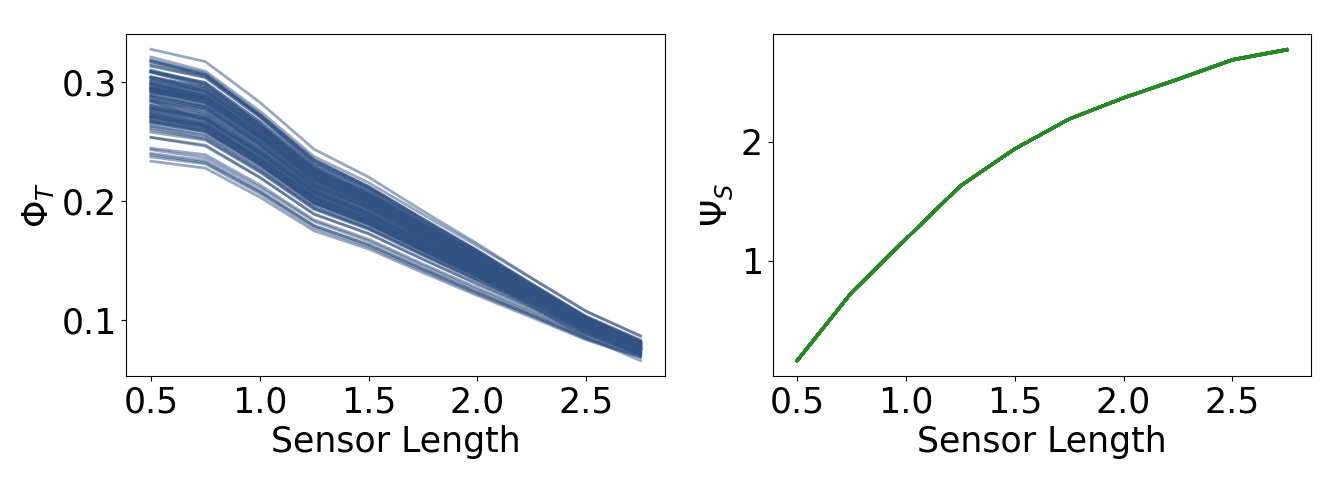}
\caption{Integrated information $\Phi_{T}$ and morphological computation $\Psi_{S}$ for the fully coupled agents.} \label{Fig8}
\end{figure}

\begin{figure}[h]
\centering
\includegraphics[width = 1\textwidth]{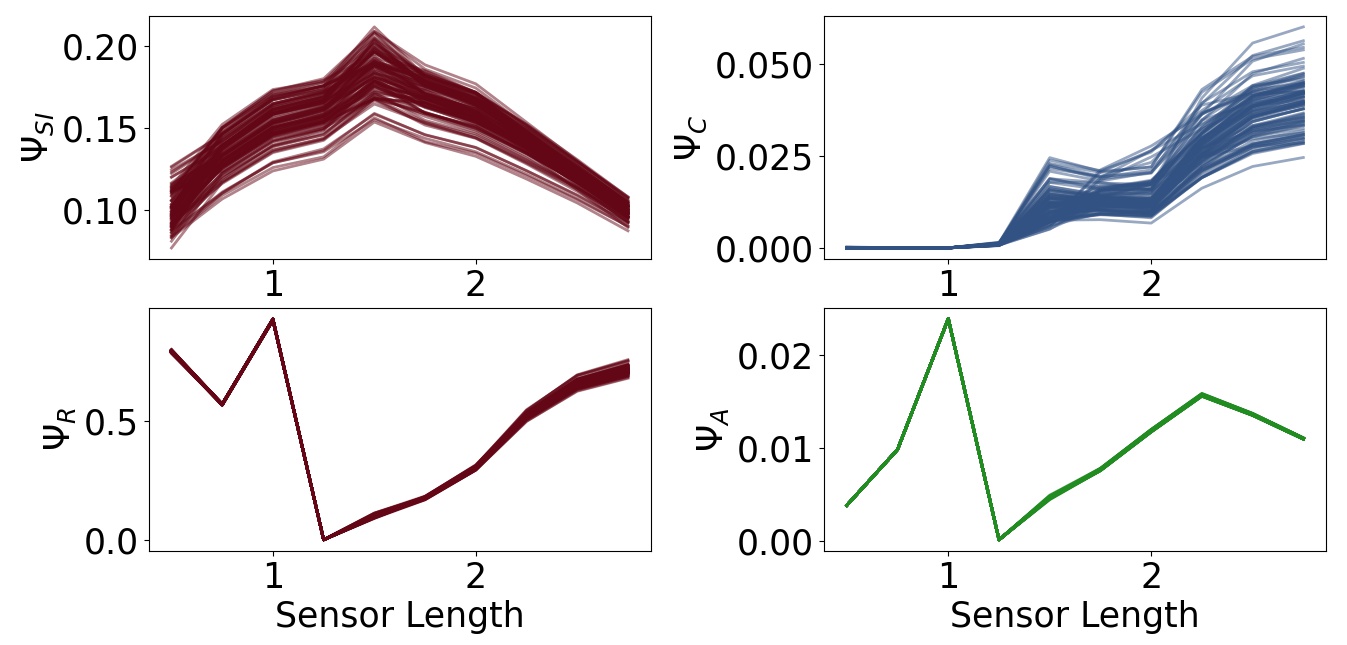}
\caption{The measures for control $\Psi_{C}$, sensory information $\Psi_{SI}$, reactive control $\Psi_{R}$ and action effect $\Psi_{A}$ for the fully coupled agents.} \label{Fig9}
\end{figure}

Now we are going to present the results for integrated information $\Phi_{T}$ and morphological computation $\Psi_{S}$, depicted in Figure \ref{Fig8}.

We observe that $\Phi_{T}$ monotonically decreases as the sensors become larger. Directly to the right, $\Psi_{S}$ exhibits the opposite dynamic. It  quantifies the influence of the past sensory input on the next sensory input given the action. Hence, taking the perspective of the agent, $\Psi_{S}$ describes the extrinsic information flow, whereas $\Phi_{T}$ only depends on the controller nodes and quantifies therefore the most intrinsic information flow. So these measures exhibit an antagonistic relationship between the outside and the inside, meaning between morphological computation and integrated information.

Note that the total information flow, $\Psi_{TIF}$, is a sum of three mutual information terms and that the first
term $I(S_{t+1} ; S_{t} , A_{t} ) $ is an upper bound of $\Psi_{S}$, the measure for morphological computation. Since  $\Psi_{S}$ is particularly high compared to the other measures, the dynamics of $I(S_{t+1} ; S_{t} , A_{t} ) $ are dominating  $\Psi_{TIF}$, leading to the monotonic increase in Figure \ref{Fig7}.

In Figure \ref{Fig9} in the first row we see the measures $\Psi_{SI}$ and $\Psi_{C}$. 
The measure $\Psi_{SI}$  quantifies how important the information flow from the sensors to the controller is. For a length below 1 the sensors are too short and above approximately 2 too long to carry information that is valuable for the controller.
 The importance of the commands sent from the controller to  the actuators is measured by $\Psi_{C}$. Between 0.5 and 1.25 this value is very close to 0, which means, that the controller has next to no influence on the behavior of the agent. In this case the sensors are so short that the agents need to react directly to it.

Hence, although $\Phi_{T}$ has its maximum values at a sensor length of 0.5, the integrated information does not have a significant impact on the behavior of the agents. Therefore the importance of the information flow in the controller of an embodied agent depends additionally on the information flowing to and from the controller.

The measure for reactive control is shown in the second row. In the case of short sensors, the information needs to get passed directly to the actuators. 
Now we will compare $\Psi_{R}$ to $\Psi_{A}$, depicted on the bottom right in Figure \ref{Fig9}. The latter one is defined as the action effect, meaning the higher $\Psi_{A}$ is, the more influence the actuators have on the next sensor state. The maximum of $\Psi_{R}$ and $\Psi_{A}$ are at a sensor length of 1, which
results in the local maximum of $\Psi_{TIF}$ in Figure \ref{Fig7}. Both graphs show a similar dynamic between sensors of length 1 to 2.25. If the sensors are too small, the information needs to pass directly to the actuators, but the actuators might not be able to assure survival and therefore  $\Psi_{R}$ is high, while  $\Psi_{A}$ is low. In the case of very long sensors, they detect a wall with a high probability, so that the next sensory state will again detect a wall regardless of the action taken. This leads to a high  $\Psi_{R}$ and a low  $\Psi_{A}$.

At a sensor length of 1.25, $\Psi_{R}$ is close to 0, as well as $\Psi_{C}$ and $\Psi_{A}$, which suggests that the algorithm converged to an optimum in which the next sensor state is not dependent on the action and the action is not dependent on the last sensor state. 

At a first glance the values of $\Psi_{A}$ and $\Psi_{C}$ seem to be insignificant compared to the other measures, but note that the relatively small amount is an expected result in these experiments. The last sensor state has a very high influence on the next sensor state and on the next action, since an agent that is not touching a wall
will most likely not touch a wall in the next step and move slowly, whereas an agent touching a wall will steer away and, depending on the length of the sensors, probably touch a wall in the next step. Nevertheless,
if $\Psi_{A}$ and $\Psi_{C}$ are not zero, then there exists an information flow and therefore an influence from the
actuators to the sensors and from the controller nodes to the actuators. Hence observing the dynamics and relating them to the other measures does lead to insights to the interplay of the different information flows.

In order to further substantiate the results of our analysis, we will now examine two subclasses of agents.
We will directly manipulating the architecture of the agents so that the influence on the actuators are limited. Hence we will gain insights on the importance of reactive control and the controller for the behavior of the agent. The first subclass contains agents that are incapable of reactive control and therefore all the information has to flow through the controller. Hence we call them controller driven agents in Section \ref{Sect:Controller}. The second class consists of agents in which the controller has no impact on the actuators. These will be called reactive control agents and discussed in Section \ref{Sect:Reactive}.

\subsection{Controller driven Agents} \label{Sect:Controller}
\begin{figure}[ht]
\centering
\includegraphics[width = 1\textwidth]{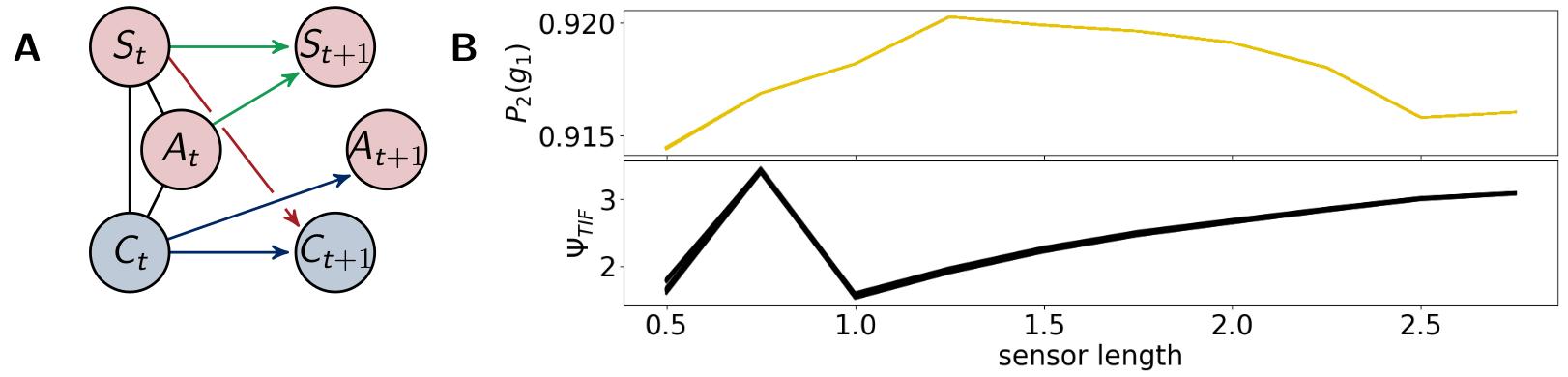} 
\caption{(A) the architecture of the controller driven agents and (B) the probability of survival (top) and
the total information flow $\Psi_{TIF}$ (bottom).}\label{Fig10}
\end{figure}

Now, we will discuss the results for the agents that are not able to use reactive control. These are displayed in Figure \ref{Fig10} on the left. We will refer to the optimized distributions by $P_{2}$.

Note that these agents are a subclass of the fully coupled ones. Hence optimizing the likelihood of success for these agents should not lead to a higher value for success than for the fully coupled agents.  But since we are using the em-algorithm that converges to local minima, we observe that controller driven agents around a sensor length 1 have a higher probability of success, as depicted on the right in Figure \ref{Fig10}. We will discuss this further in the context of Figure \ref{Fig12}. 

The results of the total information flow are similar compared to the case of the fully coupled agents after a sensor length of 1. In this case $\Psi_{TIF}$ has a global maximum at 0.75, which we will discuss in the context of Figure \ref{Fig12}.

\begin{figure}[ht]
\centering
\includegraphics[width = 0.8\textwidth]{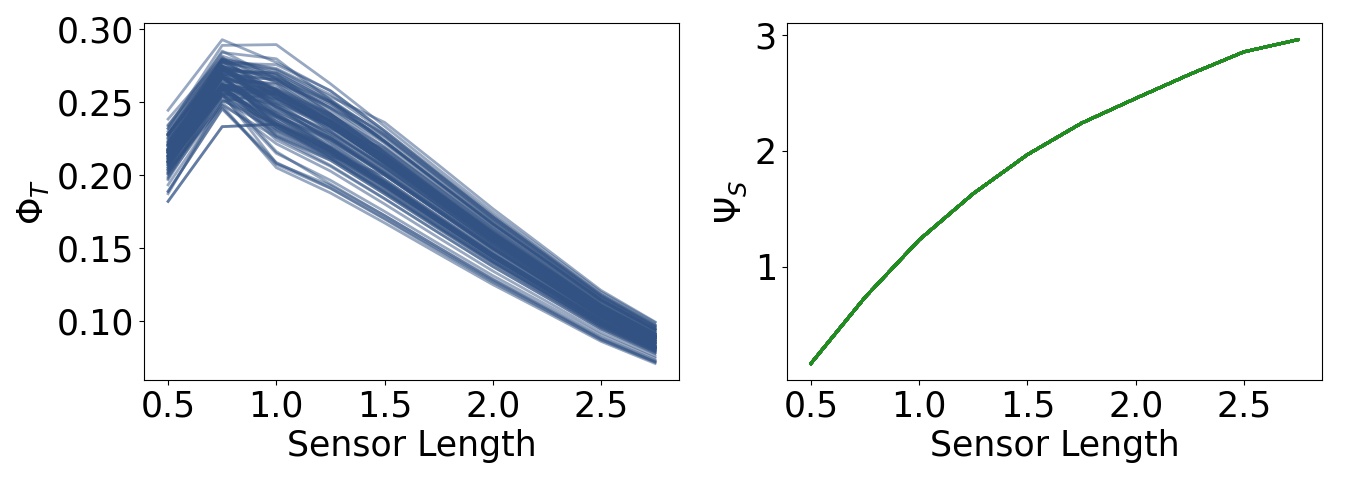}
\caption{Integrated information and morphological computation for the controller driven agents.} \label{Fig11}
\end{figure}

The measures $\Phi_{T}$ and $\Psi_{S}$ show in Figure \ref{Fig11} approximately the same values as in Figure \ref{Fig8}.  There is no change in the dynamics of $\Psi_{S}$, but $\Phi_{T}$ is lower than before at a sensor length  of 0.5. Note that $\Psi_{C}$ in Figure \ref{Fig12} is significantly higher in this case, so that the integrated information makes an impact on the actuators.

All of the measures corresponding to the controller have a spike at 0.75, at which point these agents perform better than the ones with the ability for reactive control as can be seen in the graph on the bottom left of Figure \ref{Fig12}. There the total information flow $\Psi_{TIF}$, depicted in Figure \ref{Fig10}, reaches its maximum. This spike can also be observed in $\Psi_{A}$ , meaning that the influence of the actuators on the next sensory input given the last sensory input is high. 

\begin{figure}[h]
\centering
\includegraphics[width = 1\textwidth]{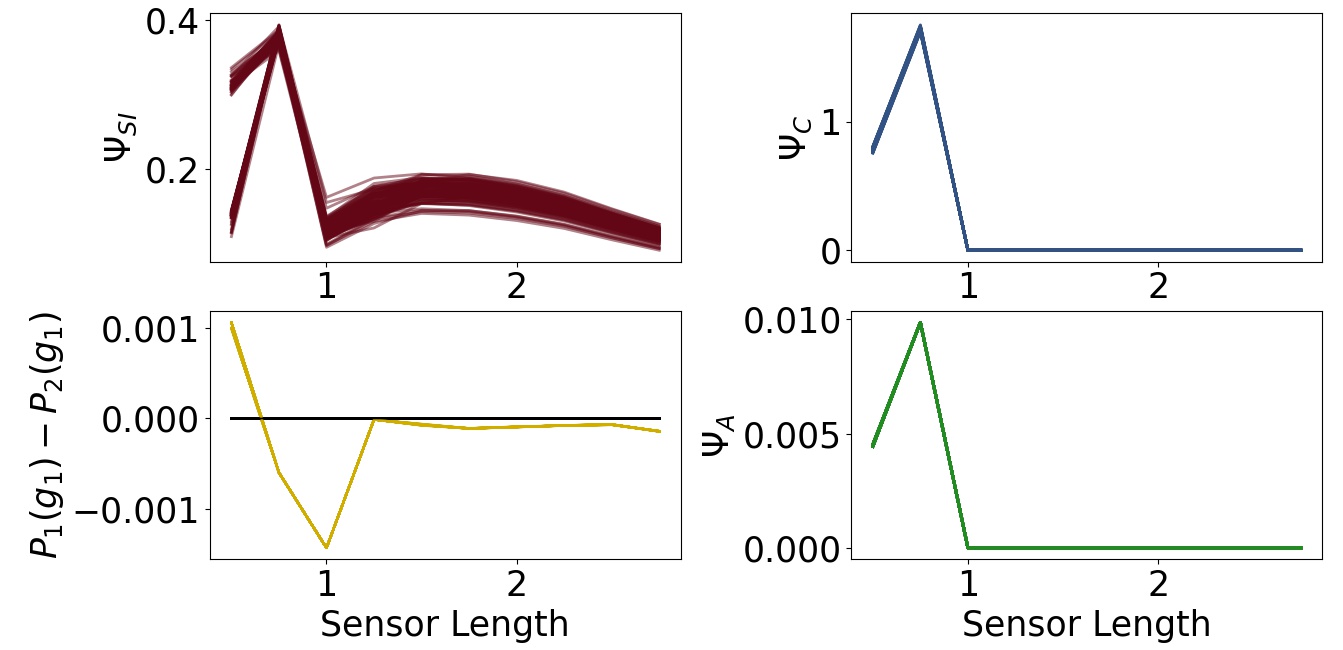}
\caption{The measures for control $\Psi_{C}$, sensory information $\Psi_{SI}$, action effect $\Psi_{A}$ for the controller driven agents and the performance difference the fully coupled agents and the reactive ones.} \label{Fig12}
\end{figure}

Additionally, looking at the goal difference depicted on the bottom left in Figure \ref{Fig12}, we see that these agents perform better than the fully coupled agents for the sensors being longer than 0.5. The black line marks the value 0. After a sensor length of 1 $\Psi_{C}$ and $\Psi_{A}$ show that the information flows from the controller to the actuators and from the actuators to the sensors are barely existent.
Therefore we come to the conclusion, that the agents converged to an optimum in which the actuators do not depend on the sensory input and have no influence on the next sensory state.  Note that $\Phi_{T}$ still shows the decreasing behavior, even though it has no impact on the actions of the agent.  

\subsection{Reactive Control Agents} \label{Sect:Reactive}

\begin{figure}[ht]
\centering
\includegraphics[width = 1\textwidth]{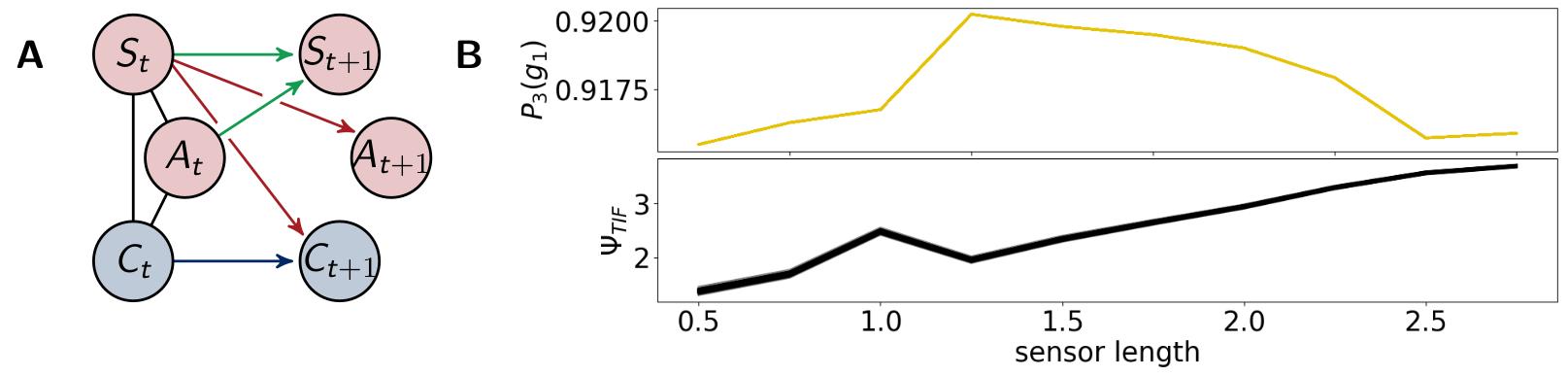}
\caption{The architecture of the reactive control agents (A) and (B) the probability of survival (top) and the total information flow $\Psi_{TIF}$ (bottom).} \label{Fig13}
\end{figure}

The architecture of the reactive control agents is shown in Figure \ref{Fig13} on the left. Here the controller has no influence on the actuators. On the right we see the probability of survival $P_{3}(g_{1})$. 

There is now significant difference between the total information flow of the fully coupled agents and the total information flow in this case.

\begin{figure}[ht]
\centering
\includegraphics[width = 0.8\textwidth]{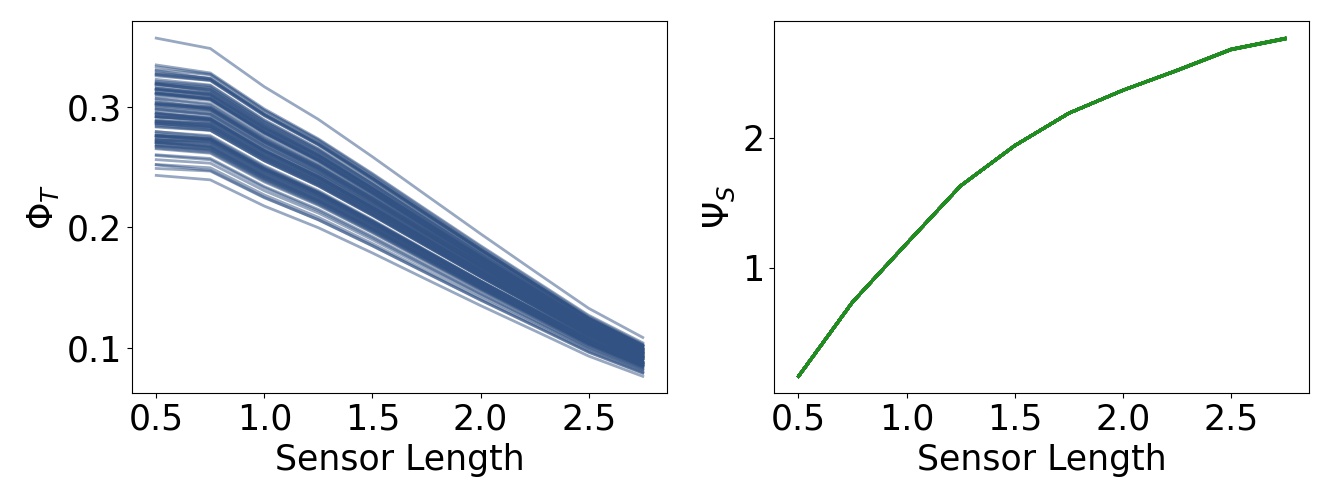}
\caption{Integrated information and morphological computation for the reactive control agents.} \label{Fig14}
\end{figure}

The measures $\Phi_{T}$ and $\Psi_{S}$ show in Figure \ref{Fig14} the same antagonistic behavior as in the fully coupled case. This demonstrates once more that only using integrated information as a measure in the case of embodied agents does not suffice if we want to understand the agent's behavior. 

\begin{figure}[ht]
\centering
\includegraphics[width = 1\textwidth]{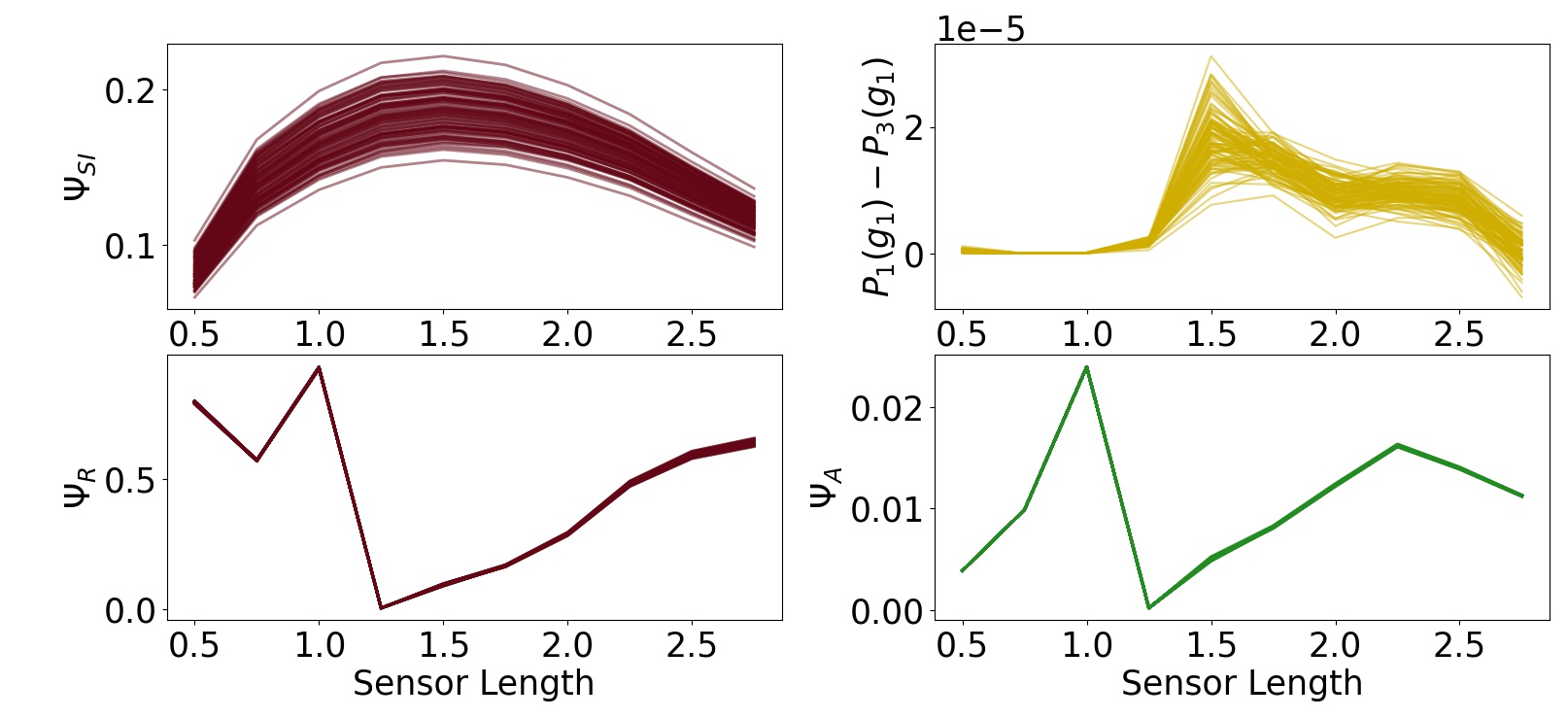}
\caption{The measures for control morphological computation $\Psi_S$, reactive control $\Psi_{R}$ and action effect $\Psi_{A}$ for the reactive control agents and the performance difference between the fully coupled agents and the reactive ones.} \label{Fig15}
\end{figure}

A closer examination of the difference in performance, depicted on the top right in Figure \ref{Fig15}, reveals that the agents connected to a controller perform better for sensors between 1.25 and 2.5. Looking back at Figure \ref{Fig9}, we see that this is approximately the region in which $\Psi_{C}$ and $\Psi_{SI}$ are both high. This supports the idea that integrated information has an impact on the behavior, when at the same time the information flows to and from the controller are high.

The other measures show the same dynamics as the corresponding measures for the fully coupled agents.

\section{Discussion}
In this article we combine different techniques in order to create a framework to analyze the information flow among an agents body, its controller and the environment. 
The main question we want to approach is how the complexity of solving a task is distributed among these different interacting parts. We demonstrate the steps in the analysis with the example of small simulated agents that are not allowed to touch the walls of a racetrack. These agents have a sufficiently simple architecture such that we are able to rigorously analyze the different information flows. Additionally, we can examine the dynamics of the information theoretic measures of an agent under changing morphological circumstances by modifying the length of the sensors.

We calculate the optimal behavior by using the concept of planning as inference, that allows us to model the conditional distributions determining the actions of the agents as latent variables. Using the information geometric em-algorithm, we are able to optimize the latent variables such that the probability of success is maximal. Here, the expectation maximization EM algorithm used in statistics is equivalent to the em-algorithm, but we chose to present the em-algorithm, because it has an intuitive geometric interpretation.  
This algorithm is guaranteed to converge, but converges to different (local) optima depending on the starting distribution. Hence this allows us to analyze various kinds of strategies that lead to a reasonably successful agent.

The distributions that are optimal regarding reaching a goal are then analyzed by applying seven information theoretic
measures. We use the measure $\Phi_{T}$ to calculate the integrated information in
the controller and we demonstrate that, although the agents have goal optimized policies, this value can be high even in cases in which it has no behavioral relevance. Therefore the importance of the information flow in the controller of an embodied agent additionally depends on the information flow to and from the controller, measured by $\Psi_{SI}$ and $\Psi_{C}$. Hence, if we want to fully understand the impact integrated information has on the behavior of an agent, it is not sufficient to only calculate an integrated information measure. This is supported by the comparison of the fully coupled agents to the reactive ones, the agents in which the controller has no impact on the actuators. It shows that the controller has a positive influence on the performance of the agents exactly in the cases in which $\Psi_{SI}$ and $\Psi_{C}$ are both high.

Comparing the morphological computation, measured by $\Psi_{S}$, to the integrated information reveals an antagonistic relationship between them. The more the agent interacts with its environment, the less information is integrated. 

The measure for reactive control $\Psi_{R}$ displays a dynamic similar to the action effect $\Psi_{A}$.  
Removing the ability to send information from the sensors directly to the actuators, in the controller driven agents, leads to agents that perform an action regardless of the sensor input for a sensor length greater than 1. 

Finally, the total information flow is an upper bound for the other measures.Therefore $\Psi_{TIF}$ combined with the measures above give us a notion of which information flow has the most influence on the system.

All in all, we present a method to completely examine the information flow among the brain, body and environment of an agent. This gives us insights into how the complexity of the task is met by the different interacting components. We observe how the morphology of the body and the architecture of the agents influence the internal information flows. The example discussed in this article is limited by its simplicity, but even in this scenario, we were able to demonstrate the value of examining the different measures. We will continue to develop these concepts further to be able to efficiently analyze more complicated agents and tasks and test them on humanoid robots. A humanoid robot can perform for example a reaching movement, which is a goal directed task that allows for more degrees of freedom and the need to integrate different information sources such as visual information and the angle of the joints.

Furthermore, we have seen in the examples presented in this
paper, that some tasks can be performed without involvement
of the controller. In contrast to the agents in this article,
which are optimized directly using planning as inference, natural
agents learn to control their body and to interact with their
environment gradually. It is intuitive to assume that learning
a new task requires much more computation in the controller
than executing an already acquired skill. Hence, it is important
to analyze the temporal dynamics of the integrated information
and morphological computation measures during the learning
process to gain insights into potential learning phases. These
different learning phases may lead us one step closer to
understanding the emergence of the senses of agency and body
ownership, two concepts closely related to the human minimal
self \cite{GallagherPaper}.

Using an agent with a more complicated morphology can lead to the opportunity to study the \enquote{degrees of freedom} problem, formulated in motor control theory. In his influential work in \cite{Bernstein} Bernstein  addresses
the difficulties resulting from the many degrees of freedom within a human body, namely the problem of choosing a particular motor action out of a number of options that lead to the same outcome. In \cite{Bernstein}, in the chapter \enquote{Conclusions towards the study of motor co-ordination}, he makes the following
observation:

\begin{quote}
\enquote{All these many sources of indeterminacy lead to the same end result; which is that the motor effect of
a central impulse cannot be decided at the centre but is decided entirely at the periphery: at the last
spinal and myoneural synapse, at the muscle, in the mechanical and anatomical change of forces in
the limb being moved, etc. }
\end{quote}

He thus emphasizes the importance of the morphology of the body for the actual movement.

There have been a number of theories further discussing this topic. In \cite{Todorov}, for
example, the authors propose a computational level theory based on stochastic optimal feedback control.
The resulting \enquote{minimum intervention principle} highlights the importance of variability in task-irrelevant
dimensions. It would be interesting to analyze, whether we observe spikes in the control value and the
integrated information that indicate a correctional motor action only for the task-relevant dimensions. 

Another theory approaching the degrees of freedom problem is the \enquote{equilibrium point hypothesis} by
Feldman and colleagues, \cite{EPH} \cite{Feldman}. There the control is modeled
by shifting equilibrium points in opposing muscles. The usage of the properties of the body in order to achieve co-ordination is directly related to the concept of morphological computation. The authors of \cite{cheapControl} study how relatively simple controllers can achieve a set of desired movements
through embodiment constraints and call this concept \enquote{cheap control}.

By applying our framework to more complex tasks, we would expect results agreeing with the observations in \cite{cheapControl}. Fewer degrees of freedom, which
are associated with strong embodiment constraints, should
lead to high morphological computation and therefore,
following the reasoning of this paper, to a small integrated
information value.

\section*{Funding}
The authors acknowledge funding by Deutsche Forschungsgemeinschaft Priority Programme “The Active Self” (SPP 2134).

\section*{Acknowledgments}
We thank Nathaniel Virgo for the design and implementation of the racetrack and for the introduction to the theory of Planning as Inference.

\section*{Appendix}
\subsection*{Optimization using the em-algorithm}
Here we are going to define the optimization using the em-algorithm.
To simplify the notation we will refer to all the sampled distributions of the visible variables by
\begin{equation}
\hat{P}_{s ,g \vert s, a} =  \tilde{P}(s_{t} \vert a_{t})   \tilde{P}(s_{t+1} \vert s_{t}, a_{t})  \tilde{P}(s_{t+2} \vert s_{t+1}, a_{t+1})\tilde{P}(g \vert s_{t+2}, s_{t+1},s_{t}, a_{t+2}, a_{t+1}, a_{t}).
\end{equation}
The em-algorithm iterates between two sets of distributions in order to find the minimal difference between them. In order to simplify the notation we will define $\mathcal{Y} = \mathcal{S} \times \mathcal{C} \times \mathcal{A}$ and $\mathcal{Z} = \mathcal{Y} \times \mathcal{Y} \times \mathcal{Y} \times \mathcal{G}$, such that $z = (s_{t},c_{t},a_{t},s_{t+1},c_{t+1},a_{t+1},s_{t+2},c_{t+2},a_{t+2}, g) \in \mathcal{Z} $. Let $\mathcal{P}(\mathcal{Z})$ be the set of probability distributions with the state space $\mathcal{Z}$ and let $\mathcal{P}^{\circ}(\mathcal{Z})$ consist of all the strictly positive distributions in  $\mathcal{P}(\mathcal{Z})$. The first set we are considering is
\begin{equation*}
\mathcal{M}_{G} := \left\lbrace Q \in \mathcal{P}(\mathcal{Z}) \vert Q(g=1)=1, \, Q(g=0)=0 \right\rbrace .
\end{equation*}
Every distribution in $\mathcal{M}_{G}$ achieves the goal with probability 1. This goal manifold is a linear family. 

The second set consists of all the distributions that factor according to the architecture of the agents, meaning that each of these distributions describes a possible behavior of an agent.
\begin{equation*}
\begin{aligned}
\mathcal{M}_{A} :=& \left\lbrace P \in \mathcal{P}^{\circ}(\mathcal{Z}) \vert P(z) = \hat{P}_{s ,g \vert s, a} \, P(c_{t} \vert a_{t}, s_{t})  P(a_{t})  \right. \\
&  \left.  \prod\limits_{i}P(a_{t+2}^{i} \vert s_{t+1}, c_{t+1}) 
\prod\limits_{i}P(a_{t+1}^{i} \vert s_{t}, c_{t}) \prod\limits_{j} P(c_{ t+1}^{j} \vert s_{t}, c_{t})  \prod\limits_{j} P(c_{t+2}^{j} \vert s_{t+1}, c_{t+1}), z \in \mathcal{Z}  \right\rbrace
\end{aligned}
\end{equation*} 
We will call $\mathcal{M}_{A}$ the agent manifold. Note that these two manifolds are disjoint, since every distribution in $\mathcal{M}_{G}$ has per definition values equal to zero and is therefore on the boundary of the probability simplex. 

The difference between elements of these two manifolds will be calculated by using the KL-divergence.

\begin{de}[KL-Divergence] \label{KL}
The Kullback-Leibler-divergence is defined as
\begin{equation*}
D(P \parallel Q) = \sum\limits_{z \in \mathcal{Z}} P(z) \, log \, \left( \dfrac{P(z)}{Q(z)} \right) 
\end{equation*}
with the conventions that $0 \cdot log \, \frac{0}{0} = 0, \, 0 \cdot log \, \frac{0}{Q(z)} = 0$ and $P(z) \cdot log \, \frac{P(z)}{0} = \infty$ for $P(z) >0$.
\end{de}
This measures how much the uncertainty of the random variable increases, if we use $Q$ instead of $P$. The KL-divergence has the following properties:
\begin{itemize}
\item[1.] $D(P \parallel Q) \geq 0$
\item[2.] $D(P \parallel Q) = 0  $ if and only if $P = Q$
\end{itemize}
Proofs of these properties can be found in \citep{Tutor} in Theorem 2.6.3. 

\begin{figure}[ht]
\centering
\includegraphics[width=0.5\textwidth]{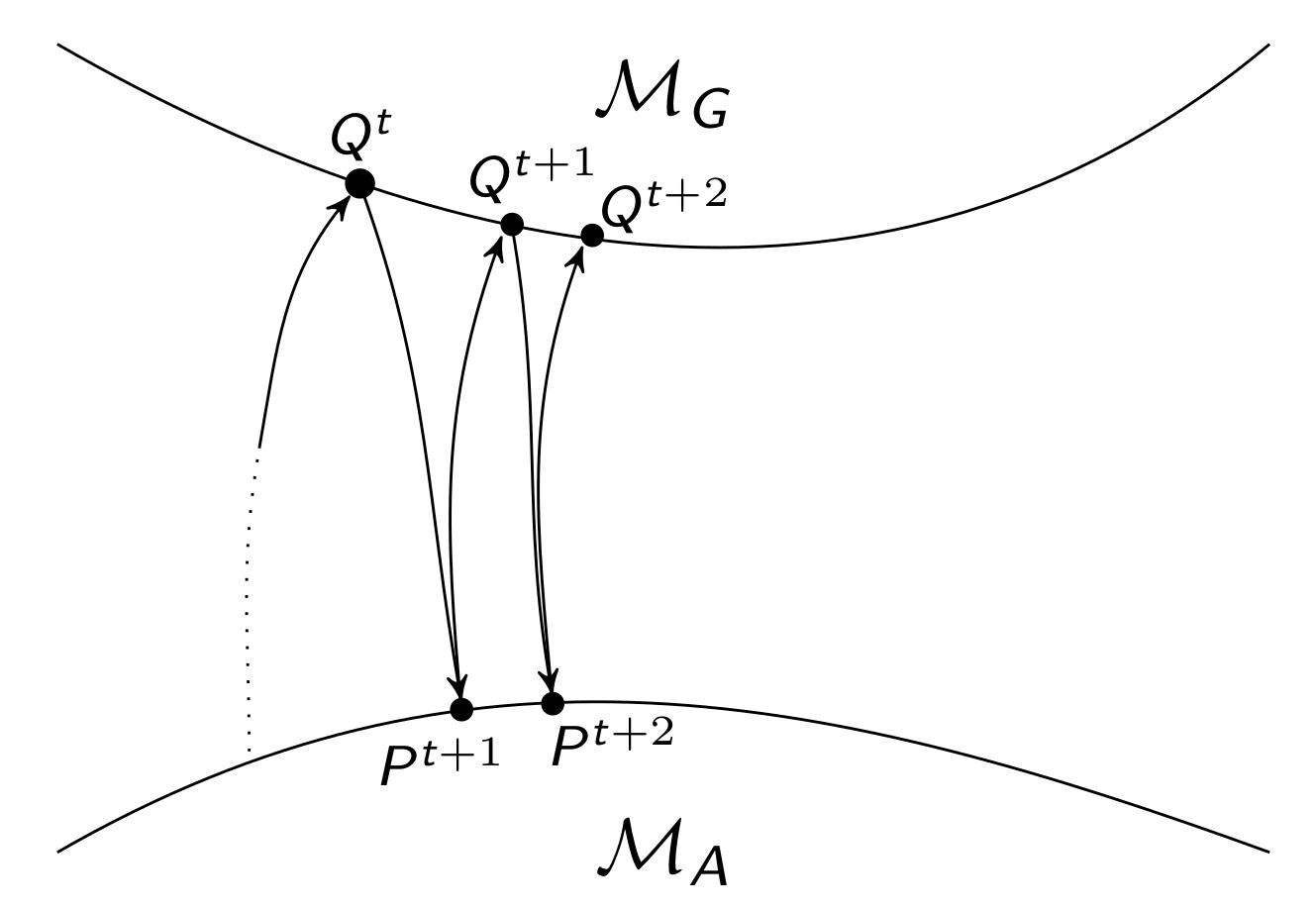}
\caption{Sketch of the em-algorithm.} \label{Fig4App}
\end{figure}

Using the em-algorithm we are able to find the minimal difference between these two manifolds 
\begin{equation*}
\inf\limits_{P \in \mathcal{M}_{A}, Q \in \mathcal{M}_{G}} D(Q \parallel P).
\end{equation*}
Therefore this procedure results in the distribution $P \in \mathcal{M}_{A}$ that is closest to achieving the goal. The algorithm works by iteratively projecting to $\mathcal{M}_{G}$ with an  $e$-projection, meaning minimizing the KL-divergence with respect to the first argument, and then projecting to $\mathcal{M}_{A}$ with an $m$-projection, defined by minimizing the KL-divergence with respect to the second argument. A sketch of this process is depicted in Figure \ref{Fig4App}.

Let $P^{0} \in \mathcal{M}_{A}$ be an arbitrary initial distribution and project this to $\mathcal{M}_{G}$ via an $e$-projection
\begin{equation*}
Q^{0} = \arginf_{Q \in \mathcal{M}_{G}} D(Q \parallel P^{0}).
\end{equation*}

Then we perform an $m$-projection to $\mathcal{M}_{A}$
\begin{equation*}
P^{1} = \arginf_{P \in \mathcal{M}_{A}} D(Q^{0} \parallel P).
\end{equation*}

Repeating this leads to
\begin{equation*}
Q^{t} = \arginf_{Q \in \mathcal{M}_{G}} D(Q \parallel P^{t}) \quad \quad  \quad \quad P^{t+1} = \arginf_{P \in \mathcal{M}_{A}} D(Q^{t} \parallel P).
\end{equation*}
This em-algorithm is guaranteed to converge, but might converge towards a local minimum, see \citep{EMvsem} or Section 5.3 of \citep{Cicar}. In our setting, the local minima are also of interest as this gives us the opportunity to not only analyze one optimal behavior, but different strategies.
 
Now it remains to show what the projections are in our case. We will start with the $e$-projection. Projecting $P^{t} \in \mathcal{M}_{A}$ to a linear family w.r.t.~the first variable is well known and can be performed in the following way
\begin{equation}
\begin{aligned}
\argmin\limits_{Q \in \mathcal{M}_{G}} D(Q \parallel P^{t}) &= Q^{t} \\
Q^{t}(z) &= P^{t}(z) \cdot \dfrac{Q(g)}{P^{t}(g)}  \label{e-proj}
\end{aligned}
\end{equation}

Note that $Q(g)$ is the same for every element in $\mathcal{M}_{G}$ and that $P \in \mathcal{M}_{A}$ is strictly positive. Therefore this expression is well defined.  

The $m$-projection can be performed as follows.
\begin{equation}
\begin{aligned}
\argmin\limits_{P \in \mathcal{M}_{A}} D(Q^{t} \parallel P) = &P^{t+1}  \\
P^{t+1}(z) =   &Q^{t}(c_{t} \vert a_{t}, s_{t}) \hat{P}(s,g \vert s, a) Q^{t}(a_{t})   
\prod\limits_{i} Q^{t}(a_{t+1}^{i} \vert s_{t}, c_{t})  \label{m-proj} \\  
&\prod\limits_{i} Q^{t}(a_{t+2}^{i} \vert s_{t+1}, c_{t+1}) \prod\limits_{j} Q^{t}(c_{ t+1}^{j} \vert s_{t}, c_{t})  \prod\limits_{j} Q^{t}(c_{t+2}^{j} \vert s_{t+1}, c_{t+1}) . 
\end{aligned}
\end{equation}

Proofs of these projections can be found in the Proofs section below.
This algorithm is equivalent to the EM-algorithm used in statistics, see Section 8.1 in \citep{Amari}  or Section 5.3 in \citep{Cicar}. 
\subsection*{Proofs}
\begin{proof}[Proof of Proposition 1]
We write $y_{t} = (s_{t},c_{t}, a_{t}) \in \mathcal{Y} = \mathcal{S} \times \mathcal{C} \times \mathcal{A} $,
\begin{equation*}
\begin{aligned}
&P(y_{t},y_{t+1}) \\
&= \sum\limits_{w_{t},w_{t+1}} P(w_{t}) \cdot P(y_{t} \vert w_{t}) \cdot P(w_{t+1} \vert w_{t}, a_{t}) \prod\limits_{k} P(s^{k}_{t+1} \vert w_{t+1}) \prod\limits_{i} P(a^{i}_{t+1} \vert s_{t},c_{t}) \prod\limits_{j} P(c^{j}_{t+1} \vert s_{t},c_{t}) \\
&= \prod\limits_{i} P(a^{i}_{t+1} \vert s_{t},c_{t}) \prod\limits_{j} P(c^{j}_{t+1} \vert s_{t},c_{t})  \sum\limits_{w_{t},w_{t+1} } P(w_{t}) \cdot P(y_{t} \vert w_{t})  \cdot P(w_{t+1} \vert w_{t}, a_{t}) \prod\limits_{k} P(s^{k}_{t+1} \vert w_{t+1}) 
\end{aligned}
\end{equation*}

The sum describes
\begin{equation*}
P(s_{t}, a_{t}, c_{t}, s_{t+1}) =  \sum\limits_{w_{t},w_{t+1} } P(w_{t}) \cdot P(y_{t} \vert w_{t})  \cdot P(w_{t+1} \vert w_{t}, a_{t}) \prod\limits_{k} P(s^{k}_{t+1} \vert w_{t+1}). 
\end{equation*}

Now we take a closer look at $P(s_{t+1} \vert s_{t}, a_{t}, c_{t})$ and show that $S_{t+1}$ is independent of $C_{t}$ given $(S_{t},A_{t})$. For that we need to describe $P(y_{t})$ in more detail. The graph corresponding to the distribution is a chain graph and therefore we are able to use Section 3.2.3 in \citep{graphModels} to gain a finer parametrization. There exist non-negative functions $f_{1}, f_{2}$, such that $P(s_{t}, a_{t}, c_{t}) = f_{2} (s_{t},a_{t},c_{t}) \sum\limits_{w_{t}} f_{1}(s_{t},w_{t})$. Using this definition results in
\begin{equation*}
\begin{aligned}
P(s_{t+1} \vert s_{t}, a_{t}, c_{t}) &= \dfrac{P(s_{t+1}, s_{t}, a_{t}, c_{t})}{P(s_{t}, a_{t}, c_{t})} \\
&=  \dfrac{f_{2} (s_{t},a_{t},c_{t}) \sum\limits_{w_{t}} f_{1}(s_{t},w_{t})\sum\limits_{w_{t+1}} P(w_{t+1} \vert w_{t},a_{t})\prod\limits_{k} P(s^{k}_{t+1} \vert w_{t+1})}{ f_{2} (s_{t},a_{t},c_{t}) \sum\limits_{w_{t}} f_{1}(s_{t},w_{t})} \\
					&=  \dfrac{\sum\limits_{w_{t}} f_{1}(s_{t},w_{t})\sum\limits_{w_{t+1}} P(w_{t+1} \vert w_{t},a_{t})\prod\limits_{k} P(s^{k}_{t+1} \vert w_{t+1})}{ \sum\limits_{w_{t}} f_{1}(s_{t},w_{t})} \\
					&=P(s_{t+1} \vert s_{t},a_{t})
\end{aligned}
\end{equation*}
Therefore the factorization of $P$ can be written as
\begin{equation*}
P(y_{t},y_{t+1})  = P(y_{t}) \cdot \prod\limits_{i} P(a^{i}_{t+1} \vert s_{t},c_{t}) \prod\limits_{j} P(c^{j}_{t+1} \vert s_{t},c_{t}) \cdot {  P(s_{t+1} \vert s_{t}, a_{t})}. \qedhere
\end{equation*}
\end{proof}

\begin{proof}[Proof of (\ref{e-proj})]
In order to proof that $Q^{t}$ is the $e$-projection of $P$ to $\mathcal{M}_{G}$ we make use of the log-sum inequality and the convention that $0 \cdot log \, 0 = 0$. Let $Q \in \mathcal{M}_{G}$ then
\begin{equation*}
\begin{aligned}
D(Q \parallel P^{t} ) &= \sum\limits_{z} Q(z) \, log \, \left(\dfrac{Q(z)}{P^{t}(z)} \right) \\
&\geq \sum\limits_{g} \left( \sum\limits_{y_{t},y_{t+1},y_{t+2}} Q(z) \right) log \, \left(\dfrac{ \sum\limits_{y_{t},y_{t+1},y_{t+2}} Q(z)}{ \sum\limits_{y_{t},y_{t+1},y_{t+2}} P^{t}(z)} \right) \\
&= 1 * log \left(\dfrac{1}{P^{t}(g)} \right) + 0* log \left( \dfrac{0}{P^{t}(g)} \right) \\
&= \sum\limits_{y_{t},y_{t+1},y_{t+2}}  P^{t}(z) \dfrac{1}{P^{t}(g)} * log \left(\dfrac{1 \cdot P^{t}(z)}{P^{t}(g) P^{t}(z)} \right) \\
 &= D(Q^{t} \parallel P^{t})
 \end{aligned}
\end{equation*}
\end{proof}

\begin{proof}[Proof of (\ref{m-proj})]
The KL-divergence between $Q^{t} \in \mathcal{M}_{A}$ and $P \in \mathcal{M}_{G}$ can be written as
\begin{equation*}
\begin{aligned}
D(Q^{t} \parallel P) &= \sum\limits_{z} Q^{t}(z) \, log \, \dfrac{Q^{t}(z)}{P(z)} \\
&= \sum\limits_{z} Q^{t}(z) \, log \, \dfrac{Q^{t}(z)}{ \hat{P}(s,g \vert s,a)} + \sum\limits_{z} Q^{t}(z) \, log \, \dfrac{1}{P(c_{t} \vert a_{t}, s_{t})} + \sum\limits_{z} Q^{t}(z) \, log \, \dfrac{1}{ P(a_{t})} \\ 
&+ \sum\limits_{z} Q^{t}(z) \, log \, \dfrac{1}{ \prod\limits_{i} P(a_{t+1}^{i} \vert s_{t}, c_{t}) } + \sum\limits_{z} Q^{t}(z) \, log \, \dfrac{1}{\prod\limits_{i} P(a_{t+2}^{i} \vert s_{t+1}, c_{t+1}) } \\
&+ \sum\limits_{z} Q^{t}(z) \, log \, \dfrac{1}{\prod\limits_{j} P(c_{t+1}^{j} \vert s_{t}, c_{t})} + \sum\limits_{z} Q^{t}(z) \, log \, \dfrac{1}{\prod\limits_{j} P(c_{t+2}^{j} \vert s_{t+1}, c_{t+1})} 
\end{aligned}.
\end{equation*}
The last six sums are called cross-entropies. It follows directly from the definition of the KL-divergence and property 1.~that the cross-entropy is greater or equal to entropy. Therefore we gain the following inequality.
\begin{equation*}
\begin{aligned}
D(Q^{t} \parallel P) \geq& \sum\limits_{z} Q^{t}(z) \, log \, \dfrac{Q^{t}(z)}{ \hat{P}(s,g \vert s,a)} + \sum\limits_{z} Q^{t}(z) \, log \, \dfrac{1}{ Q^{t}(c_{t} \vert a_{t}, s_{t})} + \sum\limits_{z} Q^{t}(z) \, log \, \dfrac{1}{  Q^{t}(a_{t})} \\ 
&+ \sum\limits_{z} Q^{t}(z) \, log \, \dfrac{1}{ \prod\limits_{i}  Q^{t}(a_{t+1}^{i} \vert s_{t}, c_{t}) } + \sum\limits_{z} Q^{t}(z) \, log \, \dfrac{1}{\prod\limits_{i}  Q^{t}(a_{t+2}^{i} \vert s_{t+1}, c_{t+1}) } \\
&+ \sum\limits_{z} Q^{t}(z) \, log \, \dfrac{1}{\prod\limits_{j}  Q^{t}(c_{t+1}^{j} \vert s_{t}, c_{t})} + \sum\limits_{z} Q^{t}(z) \, log \, \dfrac{1}{\prod\limits_{j}  Q^{t}(c_{t+2}^{j} \vert s_{t+1}, c_{t+1})} \\
&= D(Q^{t} \parallel P^{t+1})
\end{aligned}
\end{equation*}
\end{proof}
\printbibliography

\end{document}